\documentclass{degruyter-proceedings}          

\newcommand{\f}{\frac}

\newcommand{\gmr}{\frac{2GM}{c^2r}}
\newcommand{\omgmr}{\left(1-\gmr\right)}

\newcommand{\be}{\begin{equation}}
\newcommand{\ee}{\end{equation}}

\newcommand{\msun}{M_{\odot}}
\newcommand{\Msun}{M_{\odot}}
\def\pc{{\rm\thinspace pc}}
\def\Msunppc{\hbox{$\Msun\pc^{-3}\,$}}

\newcommand{\kmps}{{\rm\,km\,s^{-1}}}

\newcommand{\kg}{{\rm\,kg}}

\newcommand{\cla}{\mathcal{A}}

\def\s{{\rm\thinspace s}}
\def\erg{{\rm\thinspace erg}}
\def\ergps{\hbox{$\erg\s^{-1}\,$}}
\def\Lsun{\hbox{$\rm\thinspace L_{\odot}$}}

\def\yr{{\rm\thinspace yr}}
\def\km{{\rm\thinspace km}}

\newcommand{\rg}{r_{\rm g}}

\def\refalt@jnl#1{{\rm #1}}

\def\aj{\refalt@jnl{AJ}}                   
\def\araa{\refalt@jnl{ARA\&A}}             
\def\apj{\refalt@jnl{ApJ}}                 
\def\apjl{\refalt@jnl{ApJ}}                
\def\apjs{\refalt@jnl{ApJS}}               
\def\ao{\refalt@jnl{Appl.~Opt.}}           
\def\apss{\refalt@jnl{Ap\&SS}}             
\def\aap{\refalt@jnl{A\&A}}                
\def\aapr{\refalt@jnl{A\&A~Rev.}}          
\def\aaps{\refalt@jnl{A\&AS}}              
\def\azh{\refalt@jnl{AZh}}                 
\def\baas{\refalt@jnl{BAAS}}               
\def\jrasc{\refalt@jnl{JRASC}}             
\def\memras{\refalt@jnl{MmRAS}}            
\def\mnras{\refalt@jnl{MNRAS}}             
\def\pra{\refalt@jnl{Phys.~Rev.~A}}        
\def\prb{\refalt@jnl{Phys.~Rev.~B}}        
\def\prc{\refalt@jnl{Phys.~Rev.~C}}        
\def\prd{\refalt@jnl{Phys.~Rev.~D}}        
\def\pre{\refalt@jnl{Phys.~Rev.~E}}        
\def\prl{\refalt@jnl{Phys.~Rev.~Lett.}}    
\def\pasp{\refalt@jnl{PASP}}               
\def\pasj{\refalt@jnl{PASJ}}               
\def\qjras{\refalt@jnl{QJRAS}}             
\def\skytel{\refalt@jnl{S\&T}}             
\def\solphys{\refalt@jnl{Sol.~Phys.}}      
\def\sovast{\refalt@jnl{Soviet~Ast.}}      
\def\ssr{\refalt@jnl{Space~Sci.~Rev.}}     
\def\zap{\refalt@jnl{ZAp}}                 
\def\nat{\refalt@jnl{Nature}}              
\def\iaucirc{\refalt@jnl{IAU~Circ.}}       
\def\aplett{\refalt@jnl{Astrophys.~Lett.}} 
\def\apspr{\refalt@jnl{Astrophys.~Space~Phys.~Res.}}
\def\bain{\refalt@jnl{Bull.~Astron.~Inst.~Netherlands}}
\def\fcp{\refalt@jnl{Fund.~Cosmic~Phys.}}  
\def\gca{\refalt@jnl{Geochim.~Cosmochim.~Acta}}   
\def\grl{\refalt@jnl{Geophys.~Res.~Lett.}} 
\def\jcp{\refalt@jnl{J.~Chem.~Phys.}}      
\def\jgr{\refalt@jnl{J.~Geophys.~Res.}}    
\def\jqsrt{\refalt@jnl{J.~Quant.~Spec.~Radiat.~Transf.}}
\def\memsai{\refalt@jnl{Mem.~Soc.~Astron.~Italiana}}
\def\nphysa{\refalt@jnl{Nucl.~Phys.~A}}   
\def\physrep{\refalt@jnl{Phys.~Rep.}}   
\def\physscr{\refalt@jnl{Phys.~Scr}}   
\def\planss{\refalt@jnl{Planet.~Space~Sci.}}   
\def\procspie{\refalt@jnl{Proc.~SPIE}}   

\title{Astrophysical Black Holes}
\headlinetitle{Astrophysical Black Holes}

\lastnameone{Fabian}
\firstnameone{Andrew C.}
\nameshortone{A.\,C.~Fabian}
\addressone{Institute of Astronomy, Madingley Road, Cambridge, CB3 0HA}
\countryone{UK}
\emailone{acf@ast.cam.ac.uk}

\lastnametwo{Lasenby}
\firstnametwo{Anthony N.}
\nameshorttwo{A.\,N.~Lasenby}
\addresstwo{Kavli Institute for Cosmology, Madingley Road, Cambridge, CB3 0HA and Cavendish Laboratory, J.\,J. Thomson Avenue, Cambridge, CB3 0HE}
\countrytwo{UK}
\emailtwo{a.n.lasenby@mrao.cam.ac.uk}

\lastnamethree{}
\firstnamethree{}
\nameshortthree{}
\addressthree{}
\countrythree{}
\emailthree{}

\lastnamefour{}
\firstnamefour{}
\nameshortfour{}
\addressfour{}
\countryfour{}
\emailfour{}

\lastnamefive{}
\firstnamefive{}
\nameshortfive{}
\addressfive{}
\countryfive{}
\emailfive{}

\abstract{Black holes are a common feature of the Universe. They are observed as
stellar mass black holes spread throughout galaxies and as
supermassive objects in their centres. Observations of stars orbiting
close to the centre of our Galaxy provide detailed clear evidence for
the presence of a 4 million Solar mass black hole. Gas accreting onto
distant supermassive black holes produces the most luminous persistent
sources of radiation observed, outshining galaxies as quasars. The
energy generated by such displays may even profoundly affect the fate
of a galaxy. We briefly review the history of black holes and
relativistic astrophysics before exploring the observational evidence
for black holes and reviewing current observations including black
hole mass and spin. In parallel we outline the general relativistic
derivation of the physical properties of black holes relevant to
observation. Finally we speculate on future observations and touch on
black hole thermodynamics and the extraction of energy from rotating black holes.}

\keywords{Please insert your keywords here, separated by commas.}

\classification{Please insert AMS Mathematics Subject Classification numbers here. See \protect\url{www.ams.org/msc}}

\acknowledgments{We thank Dan Wilkins for provision of Figures~\ref{fig:accretion-disc-image} and \ref{fig:accretion-disc-extreme}, Michael Parker and Erin Kara for Fig.~\ref{fig:nustar} and Reinhard Genzel for Fig.~\ref{fig:S2-orb}. Fig.~\ref{fig:eso-gal-cen} is due to ESO/Yuri Beletsky. ANL thanks Michael Hobson and George Efstathiou as co-authors of the book \cite{2006gere.book.....H}, referred to at several points in the text above.}

\firstpage{1}

\begin{document}


\section{Introduction}

Black holes are exotic relativistic objects which are common in the Universe. It has now been realised that they play a major role in the evolution of galaxies.  Accretion of matter into them provides the power source for millions of high-energy sources spanning the entire electromagnetic spectrum. In this chapter we consider black holes from an astrophysical point of view, and highlight their astrophysical roles as well as providing details of the General Relativistic phenomena which are vital for their understanding.

To aid the reader in appreciating both aspects, we have provided two tracks through the material of this Chapter. Track 1 provides an overview of their astrophysical role and of their history within 20th and 21st century astrophysics. Track 2 (in italic text) provides the mathematical and physical details of what black holes are, and provides derivations of their properties within General Relativity. These two tracks are tied together in a way which we hope readers with a variety of astrophysical interests and persuasions will find useful.

\section{A Brief History of Astrophysical Black Holes}

Although the term ``black hole'' was coined by J.A.\ Wheeler in 1967,
the concept of a black hole is over two hundred years old. In 1783,
John Michell~\cite{michell84}  was considering how to measure the mass of a star by the
effect of its gravity on the speed of the light it emitted.  Newton had
earlier theorized that light consists of small particles. Michell
realized that if a star had the same density as the Sun yet was 500
times larger in size, then light could not escape from it.  The star
would thus be invisible.  He noted, however, that if it was orbited by
a luminous star, the measurable motion of that star would betray the
presence of the invisible one.

This prescient, but largely forgotten paper,
embodies two important concepts. The first is that Newtonian light and
gravity predicts a minimum radius $R=2GM/c^2$ for a body of mass M from
within which the body would not be visible. The second is that it can
still be detected by its gravitational influence on neighbouring
stars. The radius is now known as the Schwarzschild radius of General
Relativity and is the radius of the event horizon of a non-spinning
black hole. Black holes are now known to be common due to their
gravitational effect on nearby stars and gas. Pursuing Newtonian black
holes further leads to logical inconsistencies and also the problem
that relativity requires the speed of light to be constant.

The concept re-emerged after the publication of Eintein's General
Theory of Relativity in 1915 when Karl Schwarzschild found a solution
for a point mass.  Einstein himself "had not expected that the exact
solution to the problem could be formulated". It was not realised at
the time that the solution represented an object which would turn out
to be common in the Universe. Chandrasekhar in 1931~\cite{chandra31} discovered an
upper limit to the mass of a degenerate star and which implied the
formation of black holes (although this was not spelled out). Eddington, who wrote the first book of
General Relativity to appear in English, considered the inevitability
of complete gravitational collapse to be a {\it reductio ad absurdum} of
Chandrasekhar's formula. The concept was again ignored for a further
two decades, apart from work by Oppenheimer and Snyder~\cite{oppenheimer39} who
considered the collapse of a homogenous sphere of pressureless gas in
GR, and found that the sphere becomes cut off from communication with
the rest of the Universe. In fact, what they had discovered was the inevitability of the formation of a black hole when there is no pressure support.

With this brief introduction to the early history, we now give some details of what is now understood by a Schwarzschild black hole.

\bigskip

\hrule

{\it

\subsection{The Schwarzschild Metric}

\label{sect:s-metric}

The simplest type of black hole is described by the Schwarzschild metric. This is a vacuum solution of the Einstein field equations in the static, spherically symmetric case, and takes the form
\be
ds^2=\left(1-\frac{2GM}{c^2r}\right) dt^2
-\frac{1}{c^2}\left(1-\frac{2GM}{c^2r}\right)^{-1} dr^2 \\
-\frac{r^2}{c^2} d\theta^2 -\frac{r^2 \sin^2\theta}{c^2} d\phi^2
\label{eqn:S-metric}
\ee
where $M$ is the mass of the central object. Most of the observational properties of black holes that we need follow from this metric, rather than needing the deeper level of the field equations themselves to understand them, but to give an indication of the achievement of Schwarzschild, and since we will wish to consider briefly more general spherically-symmetric black holes as well, we give a brief sketch of how the metric is derived. For more details, and a description of the sign conventions we are using, see Chapters 8 and 9 of \cite{2006gere.book.....H}.

The field equations themselves are
\be
G_{\mu\nu} = -8 \pi T_{\mu\nu}
\ee
where the {\em Einstein} tensor $G_{\mu\nu}$ is a trace-reversed version of the {\em Ricci} tensor $R_{\mu\nu}$ and $T_{\mu\nu}$ is the matter stress-energy tensor (SET). If the matter SET is traceless, in particular if we are in a vacuum case, then clearly $G_{\mu\nu}$ must be traceless as well, and so the vacuum equations are that all components of the Ricci tensor are zero:
\be
R_{\mu\nu} = 0
\label{eqn:vac-eqn}
\ee
The Ricci tensor is in turn a contraction, $R_{\mu\nu}={R^{\lambda}}_{\mu\nu\lambda}$ of the {\em Riemann tensor}, which is defined by
\be
{R^\mu}_{\alpha\beta\gamma} =
	{\partial {\Gamma^\mu}_{\alpha\gamma} \over
		\partial x^\beta} -
		{\partial {\Gamma^\mu}_{\alpha\beta} \over
		\partial x^\gamma}
		+ {\Gamma^\mu}_{\sigma\beta}
		{\Gamma^\sigma}_{\gamma\alpha}
		-{\Gamma^\mu}_{\sigma\gamma}
		{\Gamma^\sigma}_{\beta\alpha}
\label{eqn:rie-def}
\ee
where ${\Gamma^\mu}_{\alpha\beta}$ is the {\em connection}. For a general metric $g_{\mu\nu}$, defined by
\be
ds^2 = g_{\mu\nu} dx^{\mu} dx^{\nu}
\ee
the connection is given in terms of the metric by
\be
\Gamma_{\mu\alpha\beta} = \frac{1}{2}\left({\partial g_{\mu\beta} \over \partial x^{\alpha}} + {\partial g_{\alpha \mu} \over \partial x^{\beta}} - {\partial g_{\alpha\beta} \over \partial x^{\mu}} \right)
\label{eqn:conn-from-metric}
\ee

We can thus see that solving the apparently simple equation (\ref{eqn:vac-eqn}) involves potentially formidably complicated second order partial differential equations, and it is not surprising that Einstein considered it unlikely that an exact solution for the metric around a spherical body would be found. In the current case, where we have assumed everything is static, we can attempt an ansatz for the metric which simplifies things considerably. We let
\be
ds^2 = A \, dt^2 - B \, dr^2 - r^2\left(d\theta^2+\sin^2\theta \, d\phi^2\right)
\label{eqn:gen-metric}
\ee
where $A$ and $B$ are {\em functions of $r$ alone}. This means that our PDEs become ODEs, and although the working is still complicated, we eventually find that we require
\begin{eqnarray}
R_{00} & = & -\frac{A^{\prime\prime}}{2B} +
\frac{A^\prime}{4B}\left(\frac{A^\prime}{A}+\frac{B^\prime}{B}\right)
- \frac{A^\prime}{rB}=0,  \\
R_{11} & = & \frac{A^{\prime\prime}}{2A}-\frac{A^\prime}{4A}
\left(\frac{A^\prime}{A}+\frac{B^\prime}{B}\right) -
\frac{B^\prime}{rB}=0,  \\
R_{22} & = & \frac{1}{B}-1+\frac{r}{2B} \left(\frac{A^\prime}{A}-
\frac{B^\prime}{B}\right)=0, \label{eqn:Ric-22} \\
R_{33} & = & R_{22} \sin^2\theta=0.
\end{eqnarray}
We can eliminate $A$ and find a simple equation for $B$ by forming the combination
\be
\frac{R_{00}}{A} + \frac{R_{11}}{B}+\frac{2R_{22}}{r^2}=0
\ee
which yields the ODE
\be
\frac{dB}{dr}=\frac{B\left(1-B\right)}{r}
\ee
with solution
\be
B=\frac{r}{r+C}
\ee
where $C$ is a constant. Inserting this in (\ref{eqn:Ric-22}) then yields a first order equation for $A$ of which the solution is
\be
A = \frac{a\left(r+C\right)}{r}
\ee
where $a$ is a further constant. This latter constant effectively just changes the units of time, and it is sensible to choose this so that the speed of light $c$ is 1, which we temporarily employ. We have thereby recovered the Schwarzschild metric, (\ref{eqn:S-metric}), and can identify the constant $C$ as $-2GM$.

One notices straightaway that some kind of singularity exists at $r=-C=2 GM$, but it is worth noting that this is not a singularity of the Riemann tensor (the only non-zero quantity transforming as a full tensor we have available for investigation). The entries of this all behave with $r$ like $1/r^3$, corresponding to tidal forces which make neighbouring particles move apart or together in their motion, and none have a singularity at $r=2GM$. Indeed, the form in which Schwarzschild first discovered his metric also had no singularity at this radius (see e.g.\ \cite{orig_S_grg} and the comments in \cite{2003GReGr..35..945A}), and this coupled with lack of singularity of the Riemann tensor except at the origin, perhaps contributed to the uncertainty stretching over many years as to the physical status of the distance $2GM$.

Nowadays, we of course recognise this as the ``horizon'', the point where light becomes trapped. We discuss this in more detail below, particularly in connection with the Kerr solution, but we can understand this in simple terms by looking at the `coordinate speed' of a light pulse. For such a pulse, the interval $ds=0$, and for outward radial motion this means
\be
A dt^2 - B dr^2=0, \quad i.e.\ \quad \frac{dr}{dt}=\sqrt{\frac{A}{B}}
\ee
The `horizon' is therefore where the metric coefficient $B$ goes to infinity, since this marks the point where light is no longer able to move outwards. This happens in the Schwarzschild case at $r=r_S=2GM/c^2$, which is the location of the event horizon.

Before proceeding further, we mention a few physical facts about
Schwarzschild black holes. The first is that the radius of the event
horizon, $r_S$, corresponds to $3\km$ per Solar mass. This means that the mean density
of a black hole is given by
\begin{equation}
\rho=2\times 10^{16}\left(M\over \Msun\right)^{-2} {\rm \, gm \, {cc}^{-1}}.
\end{equation}
Thus black holes with masses above $10^8\Msun$, have average densities below
that of water, or the Sun. Those above a few billion $\Msun$ have
densities below that of air. So from a mean density point of view,
supermassive black holes are not particularly dense. The
light-crossing time of the event horizon (i.e.\ a length equivalent to
its diameter) is 0.2~ms for a $10\Msun$ black hole. 20~s for
$10^6\Msun$ and about one day for $5\times 10^9\Msun$.

}

\bigskip

\hrule

\section{Relativistic Astrophysics emerges}

\label{sect:rel-astro-em}

The previous neglect of black holes changed in 1963 when two important events occurred, the
discovery of quasars by Maarten Schmidt~\cite{schmidt63} and the discovery of the
solution for a spinning black hole by Roy Kerr~\cite{Kerr:1963ud}. These combined to lead
to Relativistic Astrophysics as the combination of Astrophysics and
GR, embodied in the first Texas Symposium of Relativistic Astrophysics
in December 1963. Kerr's brief talk has in retrospect been called the
most important announcement at the Symposium but was not mentioned by
any of the three summarizers of the meeting~\cite{schucking89}.

The discovery of quasars resulted from the accurate position for the
cosmic radio source 3C273 obtained by Cyril Hazard and collaborators~\cite{hazard63}
using the lunar occultation technique. Schmidt used the 200" Hale
Telescope to take an optical spectrum of the starlike object at that
position in the Sky.  Some weeks later he identified emission lines in
the spectrum with Balmer lines of hydrogen, redshifted by
0.16. Interpreting the redshift as due to cosmic expansion means that
it is at a (luminosity) distance of 0.75 Gpc. From its apparent
brightness, the object can be inferred to be more than $10^{12}$ times
more luminous than the Sun. 3C273 can be found on optical plates taken
since the late 1800s and was soon shown to be variable, at times
changing significantly within a week. Ignoring relativistic
considerations, causality requires that significant variations cannot
occur on timescales shorter than the light crossing time of the
object.  It therefore appears that the prodigious power output of
3C273 emerges from a region of size similar to the Solar System!
\footnote{3C273 is not just a pointlike object but has an associated
linear structure which we now know to be a relativistic jet, but the
above argument is good for the main central source.}

\begin{figure}[ht]
     \centering
     \includegraphics[width=.8\textwidth]{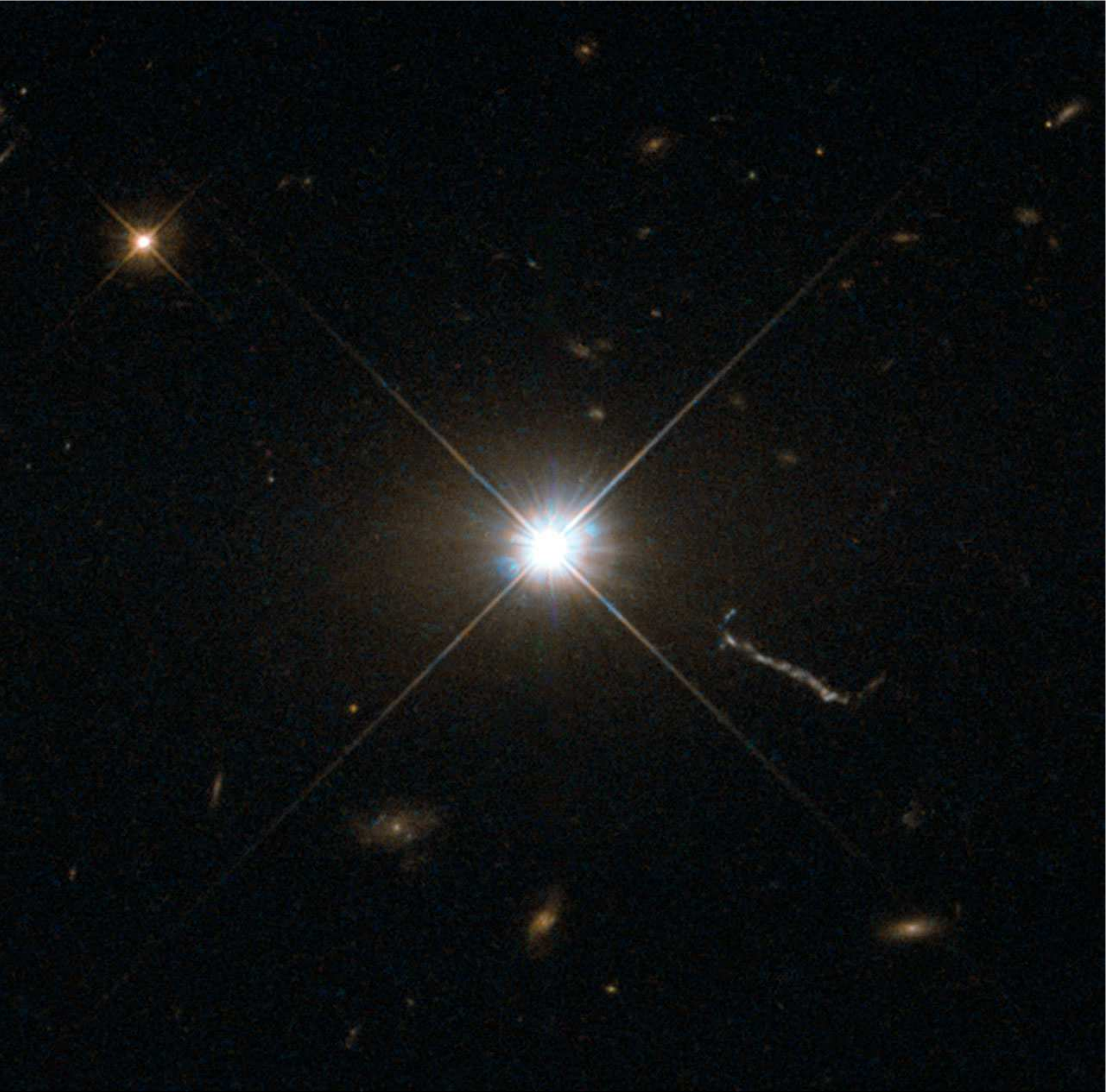}
     \caption{The first quasar identified, 3C273, as imaged by the
       Hubble Space Telescope. It lies over two billion light years
       away and is the brightest quasar in the visible Sky. The
       structure to the SW (lower right) is the outer parts of its
       relativistic jet, which is seen from radio to X-ray
       wavelengths. The projected length of the jet is over 200
       thousand light years, so its true length is nearly an order of
       magnitude larger than our galaxy. The mass of its black hole
       has been measured by the optical reverberation technique at
       almost a billion Solar masses. The X-shape is an instrumental
       artefact created by support structures in the telescope; it
       implies that the source is unresolved. 3C273 varies on week-long
       timescales indicating that the emission region is less than a
       light week across. }
\end{figure}

Kerr's solution to the Einstein field equations was not immediately
recognised as the exact solution for a spinning black hole.  We now
know the geometry described is the unique and complete description of
the external gravitational field of an uncharged stationary black hole.  It is
now thought that the power of 3C273 and other quasars is due to
accretion onto a Kerr black hole, which we now give some preliminary details of.

\bigskip

\hrule

{\it

\subsection{Rotating black holes}

It is now known that the totality of black holes can be described
using just three parameters.  Whatever makes them up, and however they
were formed, in the end only three parameters matter --- their mass,
$M$; charge, $Q$ and angular momentum $J$.  Due to the high
conductivity of interstellar matter, black holes are not likely to
have a net charge for long, so the only relevant parameter in addition
to mass is spin.\footnote{Newman et al \cite{Newman:1965my} found the
solutions to the Einstein-Maxwell field equations in 1965 and the
resulting Kerr-Newman geometry describes spinning charged black
holes.} A metric applicable to a black hole with angular
momentum was found by Kerr in 1962 \cite{Kerr:1963ud}. This metric, in the form later developed by Boyer \& Lindquist \cite{Boyer:1966qh}, is
\be
\begin{aligned}
ds^2 &=  \left(1-\frac{2GMr}{\rho c^2}\right) dt^2
- \frac{1}{c^2}\left[
\frac{-4GMra\sin^2\theta}{\rho c} dt d\phi +\frac{\rho}{\Delta} dr^2 \right.\\
&+ \left.\rho
d\theta^2 +\left(r^2+a^2+\frac{2GMra^2\sin^2\theta}{\rho c^2}\right)\sin^2
\theta d\phi^2\right],
\end{aligned}
\label{eqn:kerr-met}
\ee
where $a=(J/Mc)$ is the angular momentum of the black hole per unit mass (and
has the dimensions of distance), $\Delta=r^2-(2GMr/c^2)+a^2$ and
$\rho=r^2+a^2\cos^2\theta$. Substituting this metric into equation (\ref{eqn:conn-from-metric}) for the connection coefficients and calculating the Riemann (\ref{eqn:rie-def}) from these, the Ricci tensor $R_{\mu\nu}={R^{\lambda}}_{\mu\nu\lambda}$ will be found to vanish, and therefore satisfy the vacuum field equations
(\ref{eqn:vac-eqn}). This is a formidable calculation, however, and techniques which derive the metric in the context of more general azimuthally symmetric spacetimes (e.g.\ \cite{Doran:2002yv}) are in fact easier to carry out.

Note if the black hole is non-rotating, then
$J=a=0$ and the Kerr metric reduces to the standard Schwarzschild metric (\ref{eqn:S-metric}).
Almost certainly, all real black holes in the universe will be of the Kerr type.
The idea is that just as infalling matter will have angular momentum, so
will the material from which the black hole formed, leaving it both with a mass
and net angular momentum.

}

\bigskip

\hrule

\subsection{Black holes as energy sources}

The enormous luminosity of 3C273 and other quasars was soon linked to
black holes by Edwin Salpeter~\cite{salpeter64} and (separately) Yakov Zeldovich~\cite{zeldovich}. The energy is due to the deep gravitational potential well of the black hole which can be liberated by collisions outside the event horizon as gas falls in. If two gas clouds with small amounts of equal but opposite angular momenta fall into a Schwarzschild black hole colliding at $4r_{\rm g}$ (note $r_{\rm g}$ is defined as $GM/c^2$) and momentarily coming to rest with their kinetic energy emitted as radiation, then about 29\% of the rest mass energy of the clouds is released. More generally the angular momentum will be much larger and the radiative efficiency of accretion smaller. A
specific model involving an accretion disc was proposed by Donald
Lynden-Bell~\cite{lyndenbell69} in 1969. Due to the small size of a black hole, it is
most unlikely that gas falling into a black hole does so radially, but will
have some angular momentum which will cause it to orbit the
hole. Viscosity in the swirling gas will cause matter at smaller
radii, which is moving faster, to transfer its angular momentum
outward to material at larger radii, which is moving slower. The gas
then spreads in radius forming an accretion disc in which angular
momentum is transferred outward as matter flows inward.  The accretion
disc model of Lynden-Bell was then studied in a detail in the early
1970s by Pringle \& Rees~\cite{pringle73}, Shakura \& Sunyaev~\cite{shakura73} and, for the Kerr metric,
by Novikov \& Thorne~\cite{novikov73}. The gravitational energy liberated by the inflow
heats the disc which radiates locally as a
quasi-blackbody.  The disc is thin but may extend outward for a
considerable distance. This basic picture probably accounts for much
of the energy liberated by accreting black holes. There are important
modifications due to the magnetic nature of the ionized infalling
plasma which will be discussed later.  Accretion onto a black hole is
the most mass-to-energy efficient process known, apart from direct
matter-antimatter annihilation which, due to the rarity of antimatter
in our universe, is highly uncommon. Such accretion may account for
20-30\% of the energy released in the Universe since the recombination era.
To understand the details of how this can happen, we now examine particle motion in GR, first generally, and then applied specifically to the question of efficiency of gravitational energy release around a Schwarzschild black hole. (The more complicated issue of energy release around a Kerr black hole is treated in Section~\ref{sect:kerr-part-motion}.)

\bigskip

\hrule

{\it

\subsection{Motion in the Schwarzschild metric}

The key to most astrophysical applications of the Schwarzschild metric is how point particles and photons move in it. In General Relativity, particles move on geodesics of the metric, i.e.\ the paths with an extremal lapse of proper time (for a massive particle) or `affine parameter' (for a massless particle), along the worldline.
If we let $ds$ be the differential interval along a path, and $s_{AB}$ the total interval between two points $A$ and $B$ on a given path, then with
$\dot{x}_{\mu} \equiv dx_{\mu}/ds$ we have
\begin{equation}
\begin{aligned}
 s_{AB} &= \int_A^B ds =  \int_A^B \left[ g_{\mu\nu}
dx^{\mu}dx^{\nu}\right]^{\frac{1}{2}} =
 \int_A^B \left[ g_{\mu\nu} \frac{dx^{\mu}}{ds}\frac{dx^{\nu}}{ds}\right]^{\frac{1}{2}} \, ds\\
 &= \int_A^B G(x^{\mu},\dot{x}^{\mu}) \, ds \quad \text{where} \quad
G(x^{\mu},\dot{x}^{\mu}) = \left[ g_{\mu\nu} \dot{x}^{\mu} \dot{x}^{\nu}
\right]^{\frac{1}{2}}
\end{aligned}
\label{eqn:s-results}
\end{equation}
and finding the path which extremises $s_{AB}$ then leads to the Euler-Lagrange equations (one for each $\mu$)
\be
\frac{d}{ds}\left( \frac{\partial G}{\partial \dot{x}^{\mu}} \right) - \frac{\partial G}{\partial
x^{\mu}} =0
\label{eqn:lag-geo-eqns}
\ee
Note using the result (\ref{eqn:conn-from-metric}) is is easy to demonstrate the equivalence of (\ref{eqn:lag-geo-eqns}) with the alternative, and perhaps more familiar form
\begin{equation}
\frac{dx^\mu}{ds^2} + {\Gamma^\mu}_{\nu\sigma} \frac{dx^\nu}{ds}
\frac{dx^\sigma}{ds} = 0
\end{equation}
The advantage of the former, (\ref{eqn:lag-geo-eqns}), comes from the fact that it does not need knowledge of the connection coefficients to compute it, and also conservation laws for variables of which the metric is not an explicit function, are easy to read off. Note from its definition (compare the second and fifth quantities in the chain of equalities (\ref{eqn:s-results})), that as a numerical quantity $G(x^{\mu},\dot{x}^{\mu})$ evaluates to 1 (at least for a massive particle), which we can use to simplify formulae once its functional dependence has already been used.

So we now apply these results to the metric in the general static form given by (\ref{eqn:gen-metric}). (The advantage of carrying this out for the general form, rather than just Schwarzschild, is that it enables us to consider results for other types of black hole, such as Reissner-Nordstrom, and Schwarzschild-de Sitter --- see below.) We can assume w.l.o.g.\ that the particle motion is confined to the $\theta=\pi/2$ plane, and then the equations we find are
\be
A \dot{t}^2 - B \dot{r}^2 - r^2 \dot{\phi}^2 =1
\label{eqn:G-eq-1}
\ee
coming from $G(x^{\mu},\dot{x}^{\mu})=1$, and
\be
A \dot{t}=k, \quad \text{and} \quad r^2 \dot{\phi}=h
\label{eqn:k-and-h-defs}
\ee
coming from the Euler-Lagrange equations in $t$ and $\phi$ respectively, and where $k$ and $h$ are constants. These last two equations result from the constancy of the metric coefficients in $t$ and $\phi$ and correspond to the conservation of energy and angular momentum. For a particle of mass $m$, $k m c^2$ can be identified as the particle energy, and $h$ is the specific angular momentum per unit mass.

There is no point forming the Euler-Lagrange equation in $r$, since we already have a simple expression for $\dot{r}$ available from the combination of (\ref{eqn:G-eq-1}) and (\ref{eqn:k-and-h-defs}), yielding
\be
\dot{r}^2 = \frac{1}{B}\left(\frac{k^2}{A}-\frac{h^2}{r^2}-1\right)
\ee

We return to this general form later, but now wish to obtain results specific to the Schwarzschild case. With $A=B^{-1}=(1-2GM/r)$, and reinstating $c$ for the remainder of this section, we obtain
\be
\dot{r}^2 = c^2(k^2-1) - \f{h^2}{r^2}\omgmr +\f{2GM}{r}. \label{eq-for-rdot}
\ee

Using the usual Newtonian substitution $u\equiv 1/r$, and changing the independent variable to azimuthal angle $\phi$ rather than the interval $s$, so that it is easier to discuss the {\em shape} of the orbit, we differentiate (\ref{eq-for-rdot}) to obtain
\be
\f{d^2u}{d\phi^2} + u = \f{GM}{h^2}+\f{3GM}{c^2}u^2. \label{eq-for-u}
\end{equation}

In Newtonian gravity, the equation for planetary orbits, in the same notation,
is
\be
\f{d^2u}{d\phi^2} + u = \f{GM}{h^2},
\end{equation}
so we have {\em almost\/} got the Newtonian answer, except for the extra term
$3GMu^2/c^2$. This is what gives all the relativistic effects, and we note it
correctly goes to zero as $c\rightarrow\infty$.

\subsection{Circular orbits}

An important application of the orbit formula in the context
of high energy astrophysics, is what it tells us about {\em circular\/} orbits
in Schwarzschild geometry. These will be approximately the orbits of material
{\em accreting\/} onto black holes, since infalling material nearly always has
angular momentum, and we would not generally expect direct radial infall.

If $r$ is constant, then our equation for $u$ (\ref{eq-for-u})
yields
\be
h^2 = \f{GMr^2}{r-3GM/c^2}.
\label{eqn:eqn-for-h}
\end{equation}
Putting $\dot{r}=0$ in equation (\ref{eq-for-rdot}) gives us
\be
\frac{h^2}{c^2r^2}=\frac{k^2 r}{r-\frac{2GM}{c^2}}-1 \label{eqn-newt-tot-eng}
\end{equation}
and then putting both these last two results together
yields an equation for $k$ in
terms of $r$ alone. We derive:
\be
k = \frac{1-\frac{2GM}{rc^2}}{\sqrt{1-\frac{3GM}{rc^2}}}.
\end{equation}
Remembering $k=E/mc^2$, where $E$ is the particle energy, we have found that the
energy of a particle in a circular orbit is
\be
E_{\rm circ} = mc^2\frac{1-\frac{2GM}{rc^2}}{\sqrt{1-\frac{3GM}{rc^2}}}.\label{eqn-E-circ}
\end{equation}
An obvious check on this equation, is whether it can reproduce the Newtonian
expression for the total energy of a circular orbit in the limit of large
$r$. Using the binomial theorem we see that indeed the first two terms in an
asymptotic expansion in $r$ are
\be
E_{\rm circ} \sim mc^2 - \frac{GMm}{2r} + \ldots,
\end{equation}
Thus we get
agreement at large $r$ with the usual Newtonian expression (derived via
\be
E_{\rm tot} = \mbox{K.E.} + \mbox{P.E.} = \frac{1}{2}mv^2 - \frac{GMm}{r}=
\frac{-GMm}{2r} \quad \mbox{if} \quad \frac{mv^2}{r} = \frac{GMm}{r^2})
\end{equation}
provided
we realise that it enters as a {\em correction\/} to the rest mass energy
$mc^2$, which is the dominant term.

The equation we have just found for the energy of a circular orbit, provides us
with a remarkable amount of information about the nature of such orbits.

First we see that in the limit $m\rightarrow 0$, the orbit $r\rightarrow
3GM/c^2$ is of interest, since the singularity in the denominator can cancel
the zero at the top.  In fact this is the circular {\em photon\/} orbit at
$r=3GM/c^2$, which we will derive below in a treatment of photon motion.

Secondly, we can see which
orbits (for particles of non-zero rest mass) are {\em bound}. This will occur if $E_{\rm circ}<mc^2$,
since then we have less energy than the value for a stationary particle at
inifinity. The condition for $E_{\rm circ}=mc^2$ is that
\be
\left(1-\frac{2GM}{rc^2}\right)^2=1-\frac{3GM}{rc^2}
\end{equation}
This happens for $r=4GM/c^2$ or $r=\infty$. Thus over the range $4<r<\infty$
the circular orbits are bound.

This appears to show we can get as close as 2 Schwarzschild radii to a black
hole for particles in a circular orbit, suggesting that this is where the inner
edge of any accretion disc would be. But is such an orbit {\em stable}?
We discuss this first in the specific context of the Schwarzschild metric, and then later look at stability in the context of more general metrics of the form (\ref{eqn:gen-metric}).

\subsection{Stability of circular orbits}

In Newtonian dynamics the equation of motion of a particle in a central
potential is
\be
{1 \over 2} \left ({dr \over dt}\right )^2 + V(r) = E,
\ee
where $V(r)$ is an ``effective potential''. For an orbit around a
point mass, the effective potential is
\be
V(r) = {h^2 \over 2r^2} - {GM \over r} ,
\ee
where $h$ is the specific angular momentum of the particle. Since the $1/r^2$ term will eventually always exceed the $1/r$ term as $r \rightarrow 0$, we can see that in Newtonian dynamics a non-zero angular momentum provides an {\em angular momentum barrier}
preventing a particle reaching $r=0$ --- see Fig.~\ref{fig:newt-pot}.
\begin{figure}[ht]
\begin{center}
\includegraphics[width=0.7\textwidth]{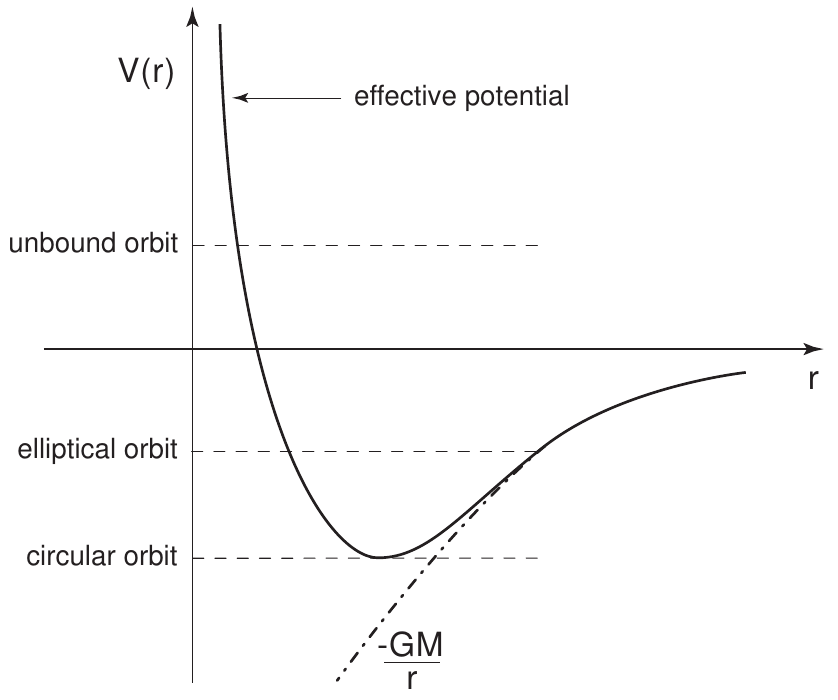}
\end{center}
\caption{The Newtonian effective potential showing how an angular
momentum barrier prevents particles reaching $r=0$.}
\label{fig:newt-pot}
\end{figure}
In this
effective potential, bound orbits have two turning points and a
circular orbit corresponds to the special case where the particle sits
at the minimum of the effective potential. However, as we have already partially seen in equation (\ref{eqn:rddot}), the same is not true in
General Relativity.

Starting with equation (\ref{eq-for-rdot}) we can rewrite this as
\be
\frac{1}{2}\dot{r}^2 + \frac{h^2}{2r^2}\left(1 - \frac{2GM}{c^2r}\right) - \frac{GM}{r} = \frac{1}{2} c^2 (k^2-1)
\ee
where we recall $k=E_{\rm part}/m c^2$ and $r^2\dot\phi=h$.

Thus, although the r.h.s. is not the particle energy here, the fact that it is constant tells us that
\be
U(r) = \frac{h^2}{2r^2}\left(1 - \frac{2GM}{c^2r}\right) - \frac{GM}{r}
\label{eqn:s-V-eff}
\ee
is an ``effective potential'' for the problem, which we can use to
study stability in the same way as in the Newtonian case. Note that
the relativistic term $(1-2GM/c^2r)$ weakens the centrifugal effect of
angular momentum at small $r$.

Differentiating this expression,
\be
{dU \over dr} =  -{h^2\over r^3} +
{3h^2 GM\over c^2 r^4}
+ {GM\over r^2},
\ee
and so the extrema of the effective potential are located at the solutions of
the quadratic equation
\be
GMr^2-h^2r+{3h^2GM\over{c^2}} = 0,
\ee
{\it i.e.} at
\be
r
= {h^2  \over {2GM}} \left \{ 1 \pm \sqrt{1- 12 \left({GM\over{h c}}\right)^2} \right
\}.
\ee
If $h = \sqrt{12}{{GM}\over c} $ {\it there is only one
extremum}, and there are no turning points in the orbit for lower
values of $h$. At this point $r=6GM/c^2=3R_{\rm s}$. Fig.~\ref{fig:U-of-r}
\begin{figure}[ht]
\begin{center}
\includegraphics[width=0.7\textwidth]{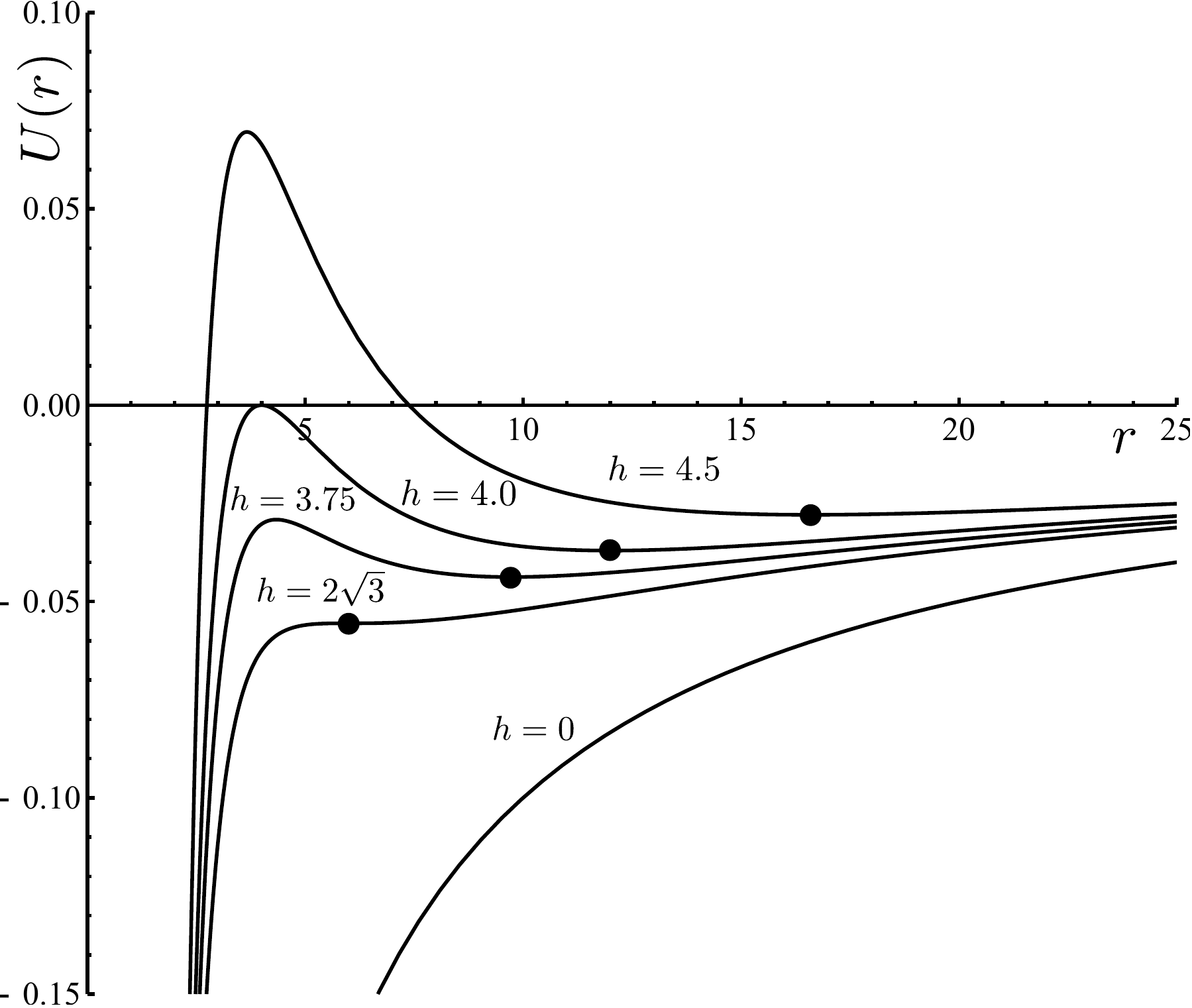}
\end{center}
\caption{The effective potential $U(r)$ plotted for several values of
the angular momentum parameter $h$ (units here have $GM/c^2=1$).}
\label{fig:U-of-r}
\end{figure}
\begin{figure}[ht]
\begin{center}
\includegraphics[width=0.7\textwidth]{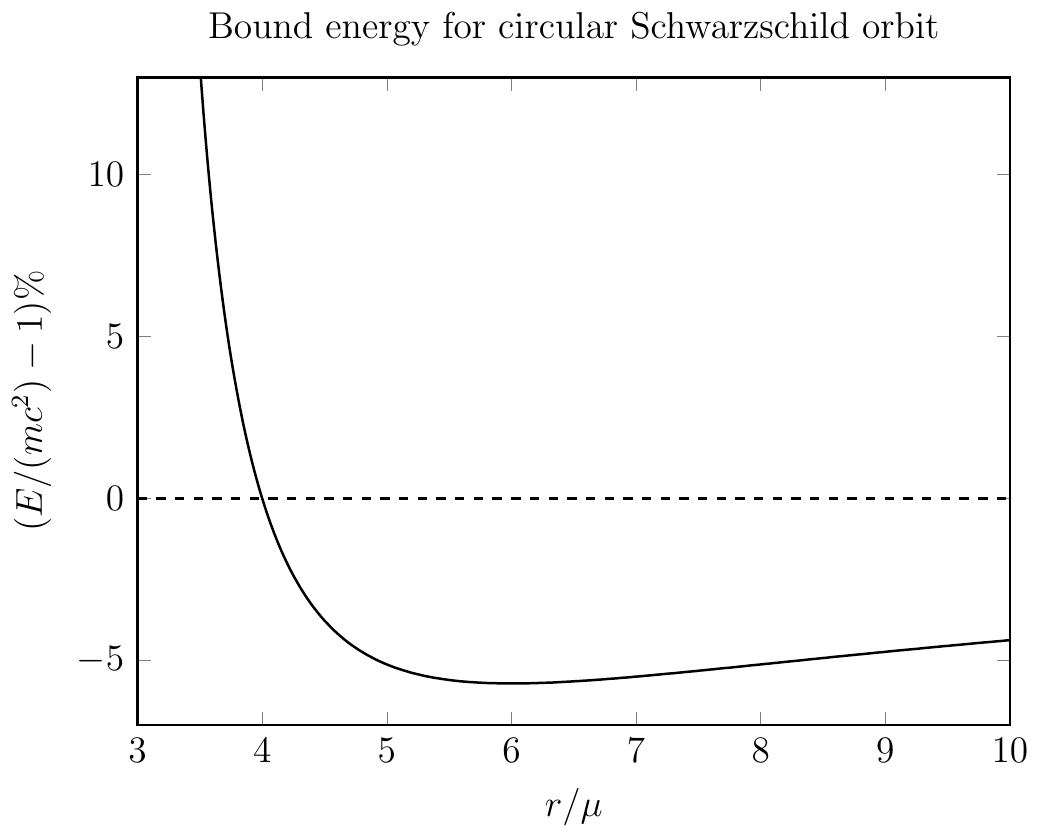}
\caption{\sl A plot of $E/(mc^2)-1$ in percent versus $r$ (the
latter measured in units of $\mu=GM/c^2$) where $E$ is
the energy of a particle of mass $m$ in a circular orbit
at radius $r$ about a Schwarzschild black hole.}
\label{fig:E-vs-r}
\end{center}
\end{figure}
shows the effective potential
for several values of $h$. The dots show the locations of
stable circular orbits. The maxima in the potential
are the locations of  {\it unstable} circular orbits.

What is the physical significance of this result? The smallest
stable circular orbit has
\be
r_{\rm min}= {6GM \over c^2}.
\ee
Gas in an accretion disc settles into circular orbits
around the compact object.
However, the gas slowly loses angular momentum because of
turbulent viscosity (the  turbulence is thought to be generated by
magnetohydrodynamic instabilities). As the gas loses angular
momentum it moves slowly towards the black hole, gaining gravitational
potential energy and heating up.
Eventually it loses enough angular momentum that it can no longer
follow a stable circular orbit and so it falls into the black hole.
On this basis, we can estimate the efficiency of energy
radiation in an accretion disc via looking at a plot
of the `fractional binding energy' $E/(mc^2)-1=k-1$
versus $r$ (see Fig.~\ref{fig:E-vs-r}).

The maximum efficiency is of order
the gravitational binding energy at the smallest stable circular orbit
divided by the rest mass energy of the gas. From the plot we can see that this will be about 6\%. More accurately, from equation (\ref{eqn-E-circ}) we see that at $r=6GM/c^2$ $k=E/(mc^2)$ is $2\sqrt{2}/3$,
hence we obtain for this efficiency
\be
\epsilon_{\rm acc} \approx 1-2\sqrt{2}/3 \approx 5.7\%
\ee
The equivalent Newtonian value, is not far away at
\be
\epsilon_{\rm acc} \approx
{1 \over 2} { G M m \over r_{\rm min}}
{1 \over mc^2} \simeq {1 \over 12} \sim 8\%.
\ee
As we will see, this value can be even larger for a black hole with spin, and an
accretion disc can convert 5-20 percent of the rest mass energy of the
gas into radiation, depending on spin. This can be compared with the
efficiency of nuclear burning of hydrogen to helium ($26$ MeV per He
nucleus),
\be
\epsilon _{\rm nuclear}  \sim 0.7 \%
\ee
Accretion discs are capable of converting rest mass energy into
radiation with an efficiency that is about 10 times greater than the
efficiency of nuclear burning of hydrogen. The `accretion power'
of black holes  causes the most energetic phenomena known
in the Universe.

}

\bigskip

\hrule

\section{Evidence from X-rays, quasars and AGN}

\begin{figure}[ht]
  \centering
  \includegraphics[width=.8\textwidth]{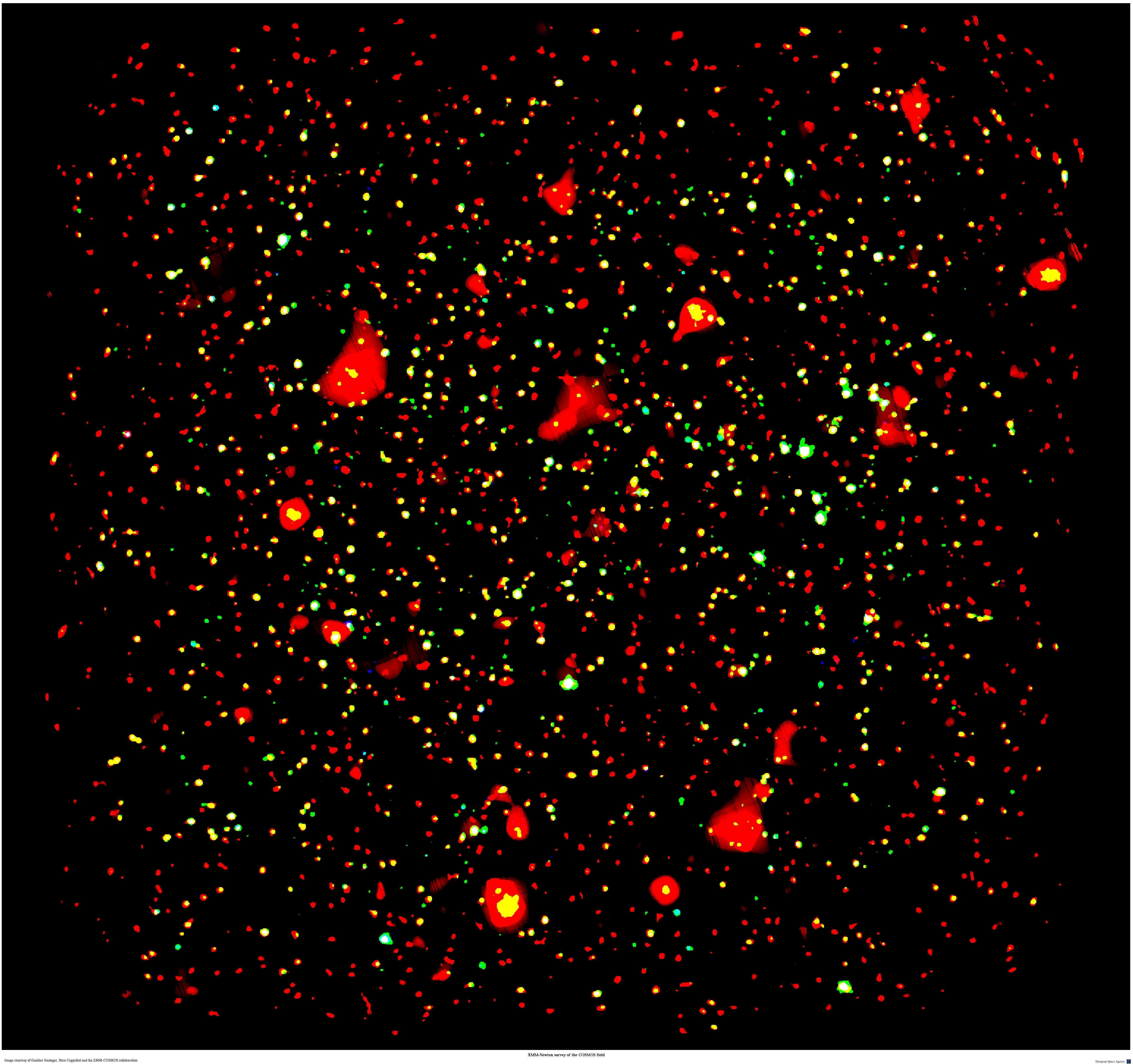}
  \caption{The XMM COSMOS field is about 30 arcmin across and shows
    the X-ray Background resolved into pointlike and extended X-ray
    sources. The pointlike sources are mostly accreting black holes
    (distant AGN) and the extended sources are due to hot gas
    pooled in the gravitational potential wells of clusters and
    groups of galaxies.}
    \label{fig:xmm-cosmos}
\end{figure}

The early 1960s also saw the birth of X-ray astronomy, when in June
1962 Riccardo Giacconi and colleagues~\cite{giacconi62} launched a sounding rocket with
Geiger counters sensitive to X-rays. As the detectors scanned across
the Sky during their 10 minutes above the Earth's atmosphere, a broad
peak of emission was discovered from the general direction of the
Scorpius constellation and a steady background of cosmic X-rays were
seen. Sco X-1 and the cosmic X-ray Background had been discovered. The
rapidly variable X-ray source Sco X-1 is a neutron star accreting
matter from an orbiting companion star and the X-ray Background is now
known to be the integrated emission from accreting supermassive black
holes at the centres of distant galaxies (see Fig.~\ref{fig:xmm-cosmos}).

The first clear association of X-ray astronomy and black holes was
made after Giacconi and colleagues studied the brightest X-ray source
in the constellation of Cygnus, Cyg X-1, using the first X-ray
astronomy satellite, Uhuru, which continuously scanned the Sky. Cyg X-1
showed chaotic variability on timescales down to a fraction of a
second. It has two relatively long-lived states, one when the spectrum is soft and
the other hard. A radio source was found to appear when it switched
into the hard state and accurate measurement of the position of that
source enabled it to be identified with a 7th magnitude B star,
HDE 226868. Separate teams of optical astronomers, Webster \& Murdin~\cite{webster72}
and Bolton~\cite{bolton72}, then found that this massive star was being swung around at $70\kmps$
by an unseen object on a 5.6 day period. This enabled mass estimates
to be made (which relied on understanding the mass of the companion B
star) yielding at least  $3.5 \Msun$ which is above the likely upper mass limit for a
neutron star of about $3\Msun$.  The small size implied by the rapid variability
combined with the lower mass limit pointed to a black hole in Cyg
X-1.

Many other black hole X-ray binary systems are now known, some of
which have much clearer mass estimates. M33 X-1, for example, in the
nearby galaxy M33 has a precise distance and stellar mass estimate
and also shows eclipses which enable the orbital inclination to be
deduced, leading to a mass for the X-ray source of $15.64 \pm 1.45 \Msun$
\cite{orosz07}.

\begin{figure}[ht]
     \centering
     \includegraphics[width=.6\textwidth]{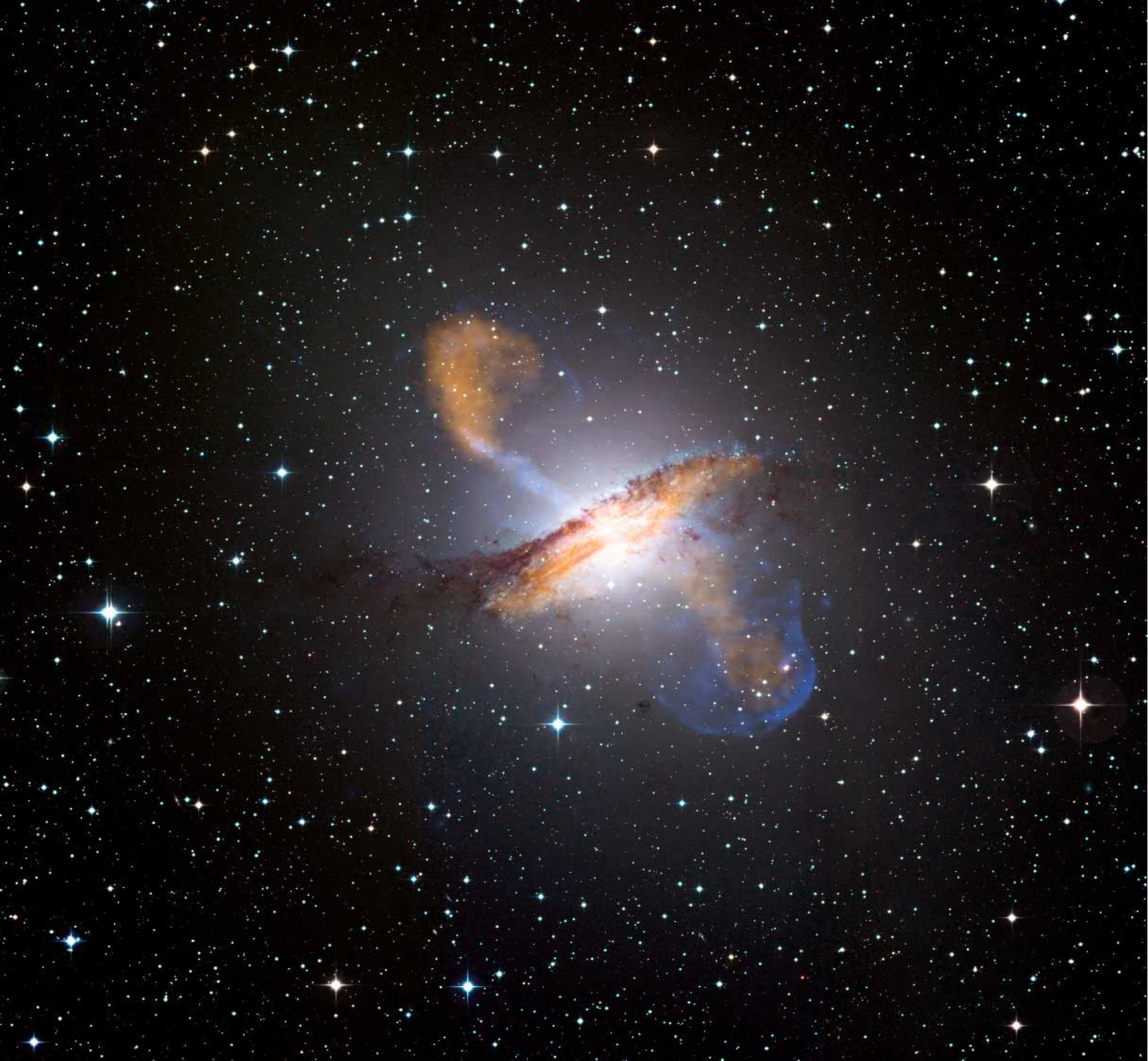}
    \includegraphics[width=.28\textwidth]{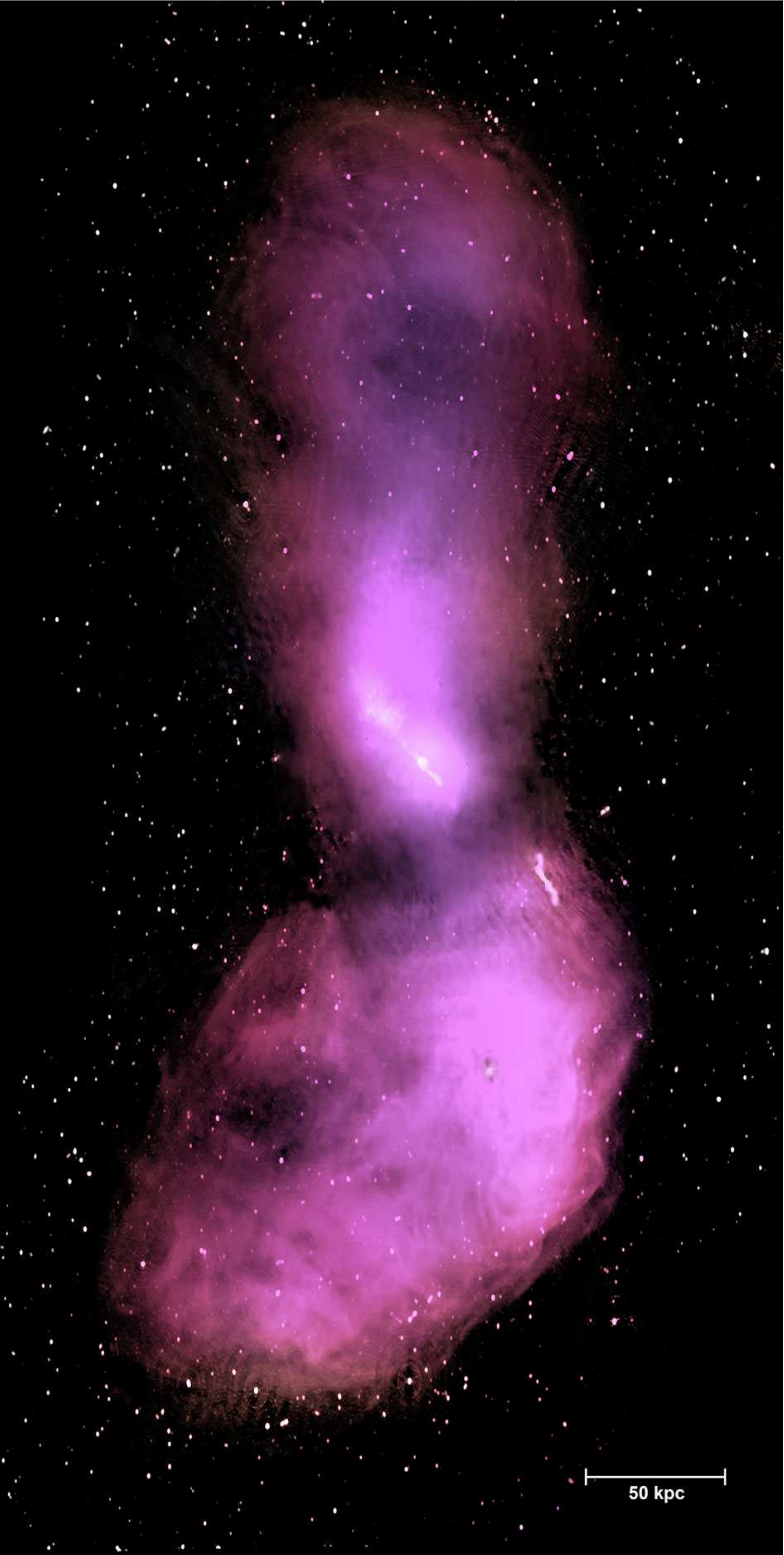}
     \caption{Centaurus A is one the nearest radio-loud AGN. The image
     on the left is a combination of infrared through X-ray images,
    its jets are seen going from the NE to SW. On the
    right is a larger scale image of low frequency radio emission
    showing the large diffuse lobes of radio-emitting plasma ejected
    by the accreting black hole out to intergalactic space.   }
    \label{fig:Cen-A}
\end{figure}

By the 1980s, many quasars and X-ray binary systems were known,
showing that black holes are common. Quasars were seen to be an
extreme part of the more general phenomenon of Active Galactic Nuclei
(AGN).  The centres of a few percent of all galaxies appear to have some
non-stellar activity in the form of bright broad emission lines, a
nonthermal radio source or an X-ray source. Quasars occur when the AGN
outshines the host galaxy in the optical band. AGN of lesser power
are known as
Seyfert galaxies or just low luminosity AGN as the power
drops. Jets of highly-collimated relativistically outflowing plasma
are found in about 10\% of AGN. An example outflow in one of the nearest AGN is shown in Fig.~\ref{fig:Cen-A}.

Unusual emission spectra at the centres of some galaxies had been known as a
phenomenon since about 1908, and the first jet was reported in M87 by
Heber Curtis in 1918~\cite{curtis18}.  Little follow-up work was done before the 1960s, apart from
Carl Seyfert's PhD thesis in the 1940s~\cite{seyfert43}.  By the end
of the 1980s most astronomers considered that AGN were powered by
accretion onto black holes, but a minority still favoured some other
explanation, such as multiple supernova outbursts. The arguments for
black holes were strong but circumstantial.  A major observational problem was that
black holes are by definition difficult to observe directly.

Quasars appear to be associated with a past phase in the evolution of
the Universe, when it was 1-5 billion years old (mostly at redshifts
of 2-4). They are now relatively uncommon compared with that
era. (3C273 is a rare quasar at low redshift: low is relative here as
it was moderately high when discovered in the early 1960s.) The
enormous powers involved mean that the accretion rates were high and
the mass doubling time of the black holes could have been a few 100
million years. Radiation pressure would have restricted the inflow to
below a limit first deduced for stars by Eddington, known as the
Eddington limit.  The limit is obtained by comparing the force of
radiation acting on an electron (through Thomson scattering of
cross-section $\sigma_{\rm T}$) at the surface of the object, with
the force of gravity on a proton there. The electron and proton,
although assumed free, are bound together by electrostatic
attraction. \be F_{\rm grav}={{GMm_{\rm p}}\over{R^2}}=F_{\rm
  rad}={L\over{4\pi R^2 h\nu}}\sigma_{\rm T} {{h\nu}\over c},\ee where
the radiative force term is the flux of photons (of typical energy
$h\nu$) times the cross-section times the momentum of a photon. This
gives \be L_{\rm Edd}={{4 \pi GMm_{\rm p}c}\over {\sigma_{\rm T}}},\ee
which is about $10^{31}$~W per Solar mass.

Since the Eddington limit is proportional to the mass of the black hole
the black hole mass can grow exponentially, if sufficient fuel
is available. If growth is at the Eddington limit, the e-folding time
of the black hole mass is about $4\times 10^7\yr$. This timescale is
sufficient for the typical massive black hole to have grown from
stellar mass (say $30\Msun$)  but it does become challenging for the most
distant billion Solar mass quasars found above redshift 6.

\section{Sgr A*, the Black Hole at the Centre of the Milky Way}

The evidence for the existence black holes changed though the 1990s
owing to careful observations of the motion of stars moving around the
centre of our own Milky Way galaxy, first starting in 1991 by Reinhard
Genzel and colleagues~\cite{genzel10} using ESO telescopes and later joined by Andre Ghez and
colleagues~\cite{ghez00} using the Keck telescopes. An interesting picture of the technique used to achieve the required image stability for this work is shown in Fig.~\ref{fig:eso-gal-cen}.
\begin{figure}[ht]
     \centering
     \includegraphics[width=.8\textwidth]{eso_guide_star_new_opt.pdf}
     \caption{The central stellar bulge of our Galaxy. A laser beam from one of the
       ESO Very Large Telescopes
       points at the very centre where Sgr A* resides. This creates an artificial star in the
     Earth's upper atmosphere which is used to stabilise images made
     of the stars orbiting the central black hole.
   }\label{fig:eso-gal-cen}
\end{figure}

\begin{figure}[ht]
     \centering
     \includegraphics[width=.8\textwidth]{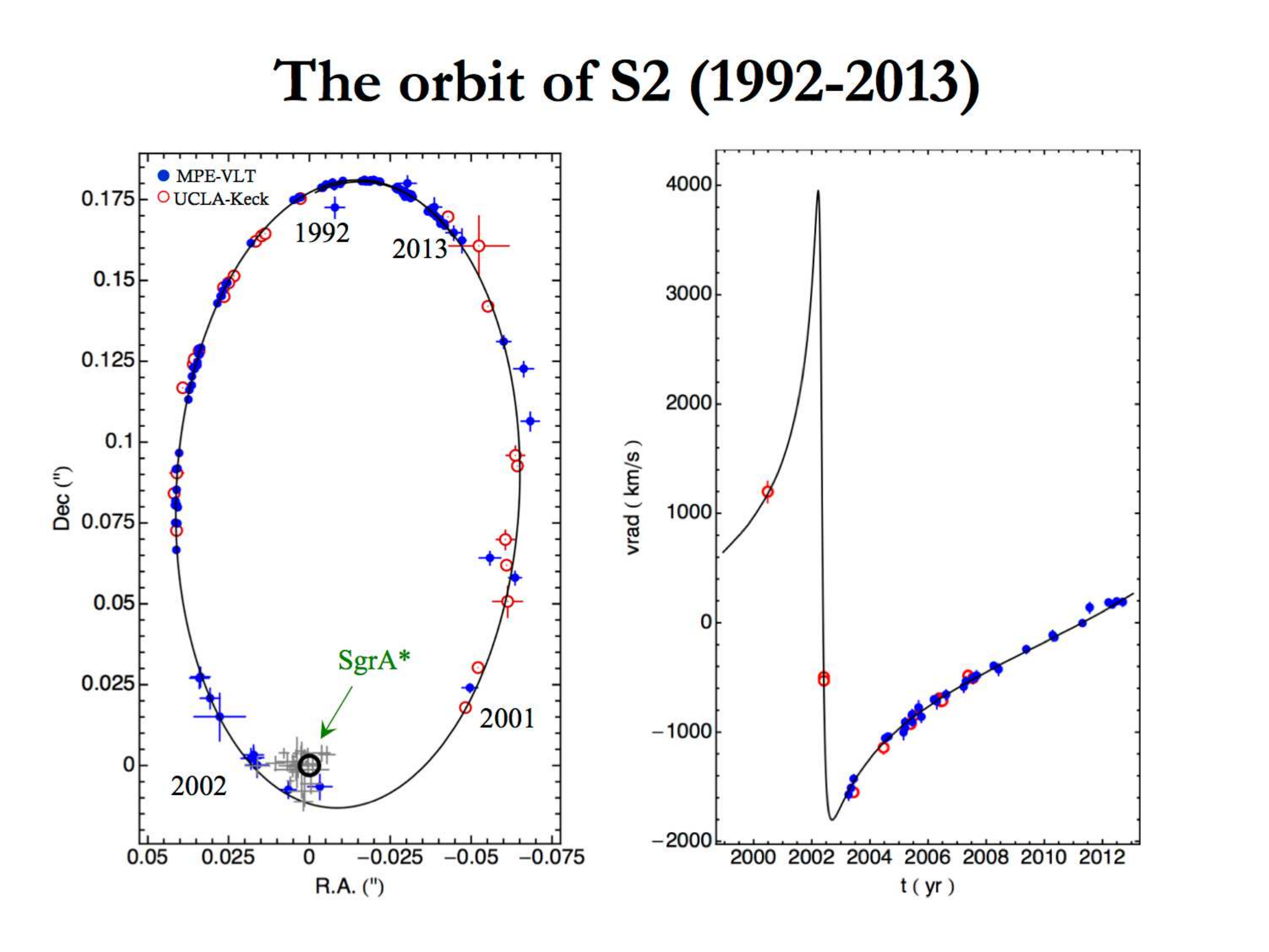}
     \caption{The path of star S2 about SgrA*. The elliptical orbit
       has a period of 15.2 years, a major axis of 5.5 light days and
       is inclined at 46 degree to the plane of the Sky. It requires
       that a mass of $4\times10^6 \Msun$ lies within a radius of 17
       light hours. The only long-lived object which is physically
       consistent is a massive black hole. }\label{fig:S2-orb}
\end{figure}

Most stars, like the Sun, orbit
the centre of the Galaxy at about $230\kmps$, but within the innermost
light year they move faster until at a distance of $2000\rg$ from the
dynamical centre a star orbits at up to $3000\kmps$.  The inferred mass
within that orbit is 4 million $\Msun$ and the dynamical centre coincides
with a strange radio source in the
constellation of Sagitarius long known as Sgr A*. It is also a  flickering infrared
and X-ray source. No stellar emission which can be attributed to say a
massive cluster of normal stars is seen.  The stellar orbits of dozens of  stars there are
known in 3D, both from their apparent motions mapped in the plane of
the Sky and from line-of-sight Doppler shifts. The orbit of one of the closest stars, S2, is shown in Fig.~\ref{fig:S2-orb}~\cite{gillesen09}.
The mass density within
the innermost orbits exceeds $10^{18}\Msunppc$ and there is nothing
known to Physics which can lie there other than a black hole. Clusters
of stars, even neutron stars, would interact and collide. The above
lower limit on the mass density of the central object is increased by
a further 5 orders of magnitude when constraints on the maximum size of Sgr A*
obtained from millimetre band Very Long Baseline Interferometry (VLBI)
are included.

The origin of the young, massive stars which are bright enough to be
tracked around the Galactic Centre is not yet fully understood. The
underlying stellar density there is so high that it should inhibit
further star formation.

As well as the issues of mass within a given radius, there is also
hope that general relativistic effects on the orbits of stars and
possibly gas clouds near Sgr A* will be observed in the near future,
so we discuss now the details of these effects, including the
important aspect of `capture' by a black hole.

\bigskip

\hrule

{\it

\subsection{GR effects on orbits}

We derived above the `shape' equation for orbits around a Schwarzschild black hole, (\ref{eq-for-u}). For the Newtonian case, then due to the harmonic kernel at the left, and the constant at the right, it is obvious that solutions of the form $u=a+b\cos\phi$ will work, i.e.\ {\em ellipses}, so if the extra term in the relativistic version (\ref{eq-for-u}) is small we expect the orbits to be modified ellipses. In the Solar System, in which a Schwarzschild metric due to the Sun applies, these modifications are very small (e.g.\ the {\em largest} effect is the 42 seconds per century precession of Mecury's orbit), but in the Schwarzschild metric around a black hole, we can expect much larger effects. For example, in Fig.~\ref{fig:peri-advance} we show the motion for an object starting about 10 Schwarzschild radii out in an almost circular orbit. The gaps between the markers on the orbit are laid down at equal intervals of time, and so indicate how fast the object is moving. We can see that strongly precessing ellipses are obtained, but with a motion that looks as though it will continue indefinitely (which it does).

However, in Fig.~\ref{fig:orbit-capture}, we show what happens when starting from the same point, and also moving tangentially, but now with a smaller specific angular momentum. This time, the object is clearly `captured' by the black hole, and ends up crossing the black hole horizon.
\begin{figure}[ht]
\begin{center}
\includegraphics[width=0.7\textwidth]{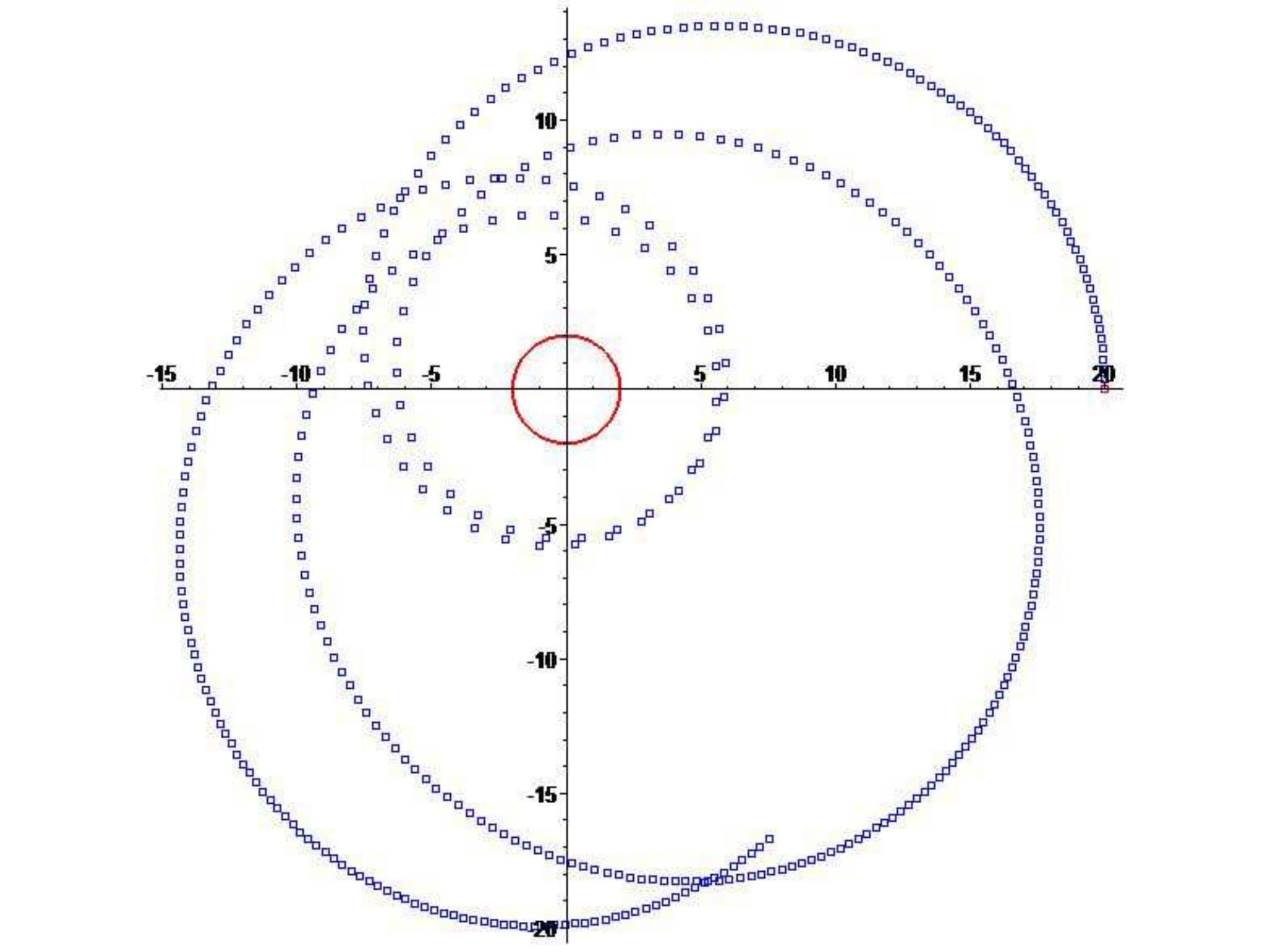}
\caption{\small Advance of perihelion in a Schwarzschild metric. The
units of distance are $GM/c^2$, where $M$ is the mass of the central body. The
particle is launched tangentially and given a specific angular momentum $h$ of 3.75 where the circular orbit $h$ would be $4.85$. The circle marks the Schwarzschild radius of the central black hole.}
\label{fig:peri-advance}
\end{center}
\end{figure}
\begin{figure}[ht]
\begin{center}
\includegraphics[width=0.775\textwidth]{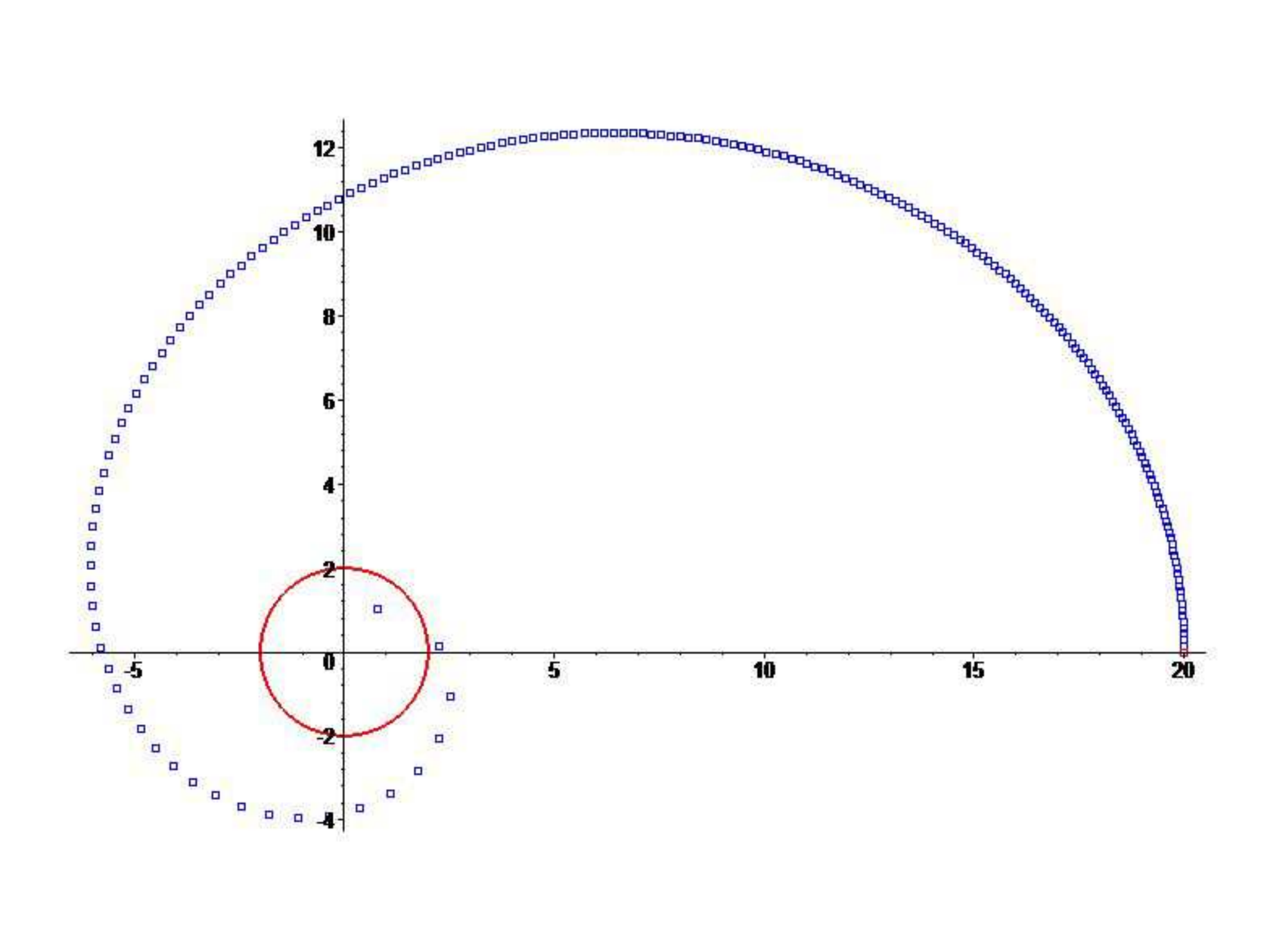}
\caption{\small As for preceding figure but now for a particle projected with $h=3.5$ (in units
with $GM/c^2=1$).}
\label{fig:orbit-capture}
\end{center}
\end{figure}

We can gain some insight into how the `capturing' comes about, by considering the orbit equation in a different form. For a massive particle, the interval $s$ we have been using is just the {\em proper time} of the particle, which we label $\tau$. If we start with equation (\ref{eq-for-rdot}), differentiate with respect to interval $\tau$ and
then remove first derivatives, $dr/d\tau$ using the original equation again,
one arrives at the following:
\be
\frac{d^2r}{d\tau^2} = -\frac{GM}{r^2} +\frac{h^2}{r^3} -
\frac{3h^2GM}{c^2r^4}.
\label{eqn:rddot}
\end{equation}
The first two terms are very Newtonian-like, and correspond to an inward
gravitational force and a repulsive term, proportional to angular momentum
squared, which is basically the `centrifugal force'. What is new is the third
term, also proportional to angular momentum squared, but this time acting
inwards. This shows that close to the hole, specifically within the radius
$r=3GM/c^2$, centrifugal force `changes sign', and is directed inwards, thus
hastening the demise of any particle that strays too close to the hole, as in the example of Fig.~\ref{fig:orbit-capture}.

\subsubsection{Orbital precession}

The equation we wish to solve is
\be
\frac{d^2u}{d\phi^2}+u = \frac{GM}{h^2} +\frac{3GM}{c^2}u^2
\quad \quad (u\equiv \frac{1}{r}) \label{eqn-u}
\end{equation}
in the limit that the departure from Newtonian motion is small. This would
apply to the motion of the planets in our solar system for example.

The Newtonian solution to this equation is
\be
u = \frac{GM}{h^2}(1+e\cos\phi),
\end{equation}
and so we can use this as a first approximation, and then iterate to get
a better one. Substituting into the r.h.s. of (\ref{eqn-u}), we obtain the new equation
\be
\frac{d^2u}{d\phi^2}+u = \frac{GM}{h^2}
+\frac{3(GM)^3}{c^2h^4}(1+e\cos\phi)^2.
\end{equation}
This will be solved by $u = \frac{GM}{h^2}(1+e\cos\phi)$ + the
particular integral (P.I.) of the equation:
\be
\frac{d^2u}{d\phi^2}+u = A(1+2e\cos\phi+e^2\cos^2\phi)
\end{equation}
where $A=3(GM)^3/c^2h^4$ is very small. The P.I.\ can be found to be:
\be
A\left(1+e\phi\sin\phi+e^2\left(\frac{1}{2}-\frac{1}{6}\cos 2\phi\right)\right).
\end{equation}
Now, in this expression, the first and third terms are tiny, since $A$ is.
However, the second term, $Ae\phi\sin\phi$, might be tiny at first, but will
gradually grow with time, since the $\phi$ part (without a $\cos$ or $\sin$
enclosing it) means it is {\em cumulative}. We must
therefore retain it, and our second approximation is
\be
u = \frac{GM}{h^2}(1+e\cos\phi+\delta e \phi\sin\phi)
\end{equation}
where
\be
\delta=\frac{3(GM)^2}{h^2c^2} \ll 1.
\end{equation}
Using
\be
\cos\left(\phi(1-\delta)\right) = \cos\phi\cos\delta\phi +
\sin\phi\sin\delta\phi \approx\cos\phi+\sin\phi \, \delta\phi \quad
\mbox{for} \quad \delta\ll 1,
\end{equation}
we can therefore write
\be
u\approx \frac{GM}{h^2}\left(1+e\cos[\phi(1-\delta)]\right).
\end{equation}
$u$ is therefore periodic, but with period $\frac{2\pi}{1-\delta}$. The $r$ values
thus repeat on a cycle which is slightly larger than $2\pi$,
%
%
and we find
\be
\Delta\phi = \frac{2\pi}{1-\delta}-2\pi \approx 2\pi\delta =
\frac{6\pi(GM)^2}{h^2c^2.}
\end{equation}

But from the geometry of the ellipse, and the Newtonian solution,
%
%
where we know $l=h^2/GM$ and $l=a(1-e^2)$, we can get the final result:
\be
\Delta\phi = \frac{6\pi(GM)^2}{c^2l(GM)}=\frac{6\pi GM}{a(1-e^2)c^2}.
\end{equation}
For example, Mercury's orbit has $a=5.8\times10^{10}{\rm\, m}$, eccentricity
$e=0.2$ and we know $\msun=2\times10^{30}\kg$. Therefore our prediction for the
precession is
\begin{eqnarray*}
\Delta\phi &=& 5\times 10^{-7} \quad \mbox{radians per orbit} \\
           &=& 0''.1 \quad \mbox{per orbit}.
\end{eqnarray*}
Since Mercury's orbital period is 88 days, we would thus expect to accumulate a
precession of $43''$ per century. This is what is observed after correction for
the perturbations due to the other planets, which cause a {\em total\/}
precession of more like $5000''$ per century.

Turning now to the Galactic Centre, the star S2, whose orbit was shown in Fig.~\ref{fig:S2-orb}, has an eccentricity of 0.876 and semi-major axis of 980 AU \cite{2005ApJ...628..246E}. Taking the mass of the Galactic Centre black hole as $4\times 10^6\msun$, then our formula yields a precession of 11 arcmin per orbit, which takes about 15 years. This sounds large, but projected on the sky at perihelion amounts to only about 0.5 milliarcsec. By the use of near-infra red interferometry, this may be observable in the relatively near future (see Section~\ref{sect:future-obs}), but as for the solar system it is competing effects from other nearby masses, and the general matter distribution near the centre, that will be key to determining whether it is the GR effect itself which is being seen \cite{2012RAA....12..995M}.

}

\bigskip

\hrule

\section{AGN Feedback}

\begin{figure}[ht]
     \centering
     \includegraphics[width=.8\textwidth]{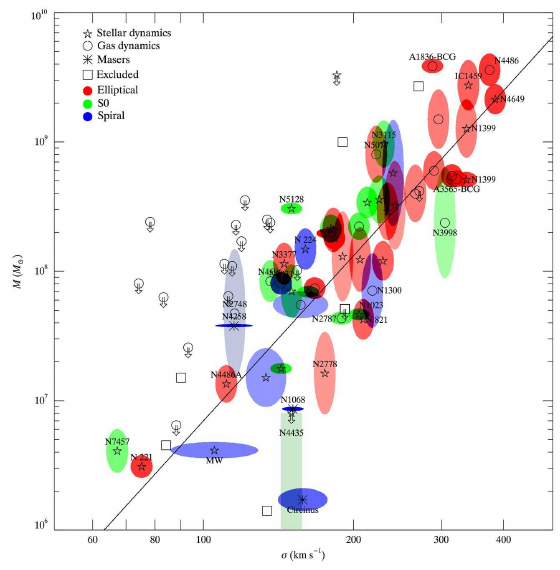}
     \caption{The black holes mass -- stellar velocity dispersion
       relation $M_{bh}-\sigma$~\cite{gultekin09}).}
       \label{fig:gultekin}
\end{figure}

It was earlier unclear whether galaxies with central black holes were
special in some way or not. Quasars could be long lived in just a few
galaxies or short lived and occur in every galaxy. Then careful
imaging of the centres of nearby galaxies with HST and other
telescopes revealed through the 1990s that most galaxies host a
massive black hole. The mass of the black hole is correlated with the
mass of the galaxy bulge\footnote{This means that the mass of any
  galactic disc  is ignored.} that it is embedded in (the Magorrian
relation~\cite{magorrian98}). Current results indicate that the host bulge has a mass 500
times that of the black hole.  Some argue that the correlation is
better if the velocity dispersion of the surroundings stars (beyond
the radius where the black hole's mass dominates) is used (the $M_{BH} -
\sigma$ relation --- an example of this is shown in Fig.~\ref{fig:gultekin}).

Both correlations show considerable scatter but
hold over several orders of magnitude.

The $M-\sigma$ relation poses the question of how the black hole at
the centre of a galaxy
knows the total mass of the host galaxy? The ratio of the physical size of the black hole
event horizon to the galaxy can be 100 million or more (a similar
scaled comparison is the Earth and a football). Some astronomers argue
that larger galaxies just have larger black holes, but the
correlations look better then that. An exciting possibility is that
the question should be turned round and it is the black hole
which controls the mass of the host galaxy. In 1998 Silk \& Rees~\cite{silk98}
proposed that it was the phenomenal energy output in the quasar phase,
as the black hole grew in mass by accretion, which was
responsible. Both grow from gas in the galaxy --- the galaxy forming
stars from the gas, the black hole accreting the gas. Eventually the
black hole is so big and powerful that it blasts all the gas from the
galaxy so both stop growing and we are left with the observed
relation.

There is clear evidence that massive black holes have grown by
accretion from the Soltan relation~\cite{soltan82}:
\begin{equation}
{\cal E}_{\rm acc}(1+z)={\eta \over(1-\eta)} \rho_{\rm BH} c^2,
\end{equation}
where $z$ is the mean redshift at which the accretion occurs.  It is a
recasting of the famous equation, $E=mc^2$, in terms of densities. The
factor of $1+z$ is due to the redshifting of the energy of the radiation;
there is no such factor for mass.  $\eta$ is the radiative efficiency
of the accretion process $L=\eta \dot M c^2$. The energy density of radiation
from accreting black holes, ${\cal E}_{\rm acc}$ can be measured from
the summed spectra of AGN (and the X-ray Background) and $\rho_{\rm
  BH}$ can be estimated from the mass function of galaxies together
with the $M_{\rm BH}-M_{\rm gal}$ relation. Results show agreement if
$\eta$ is about 0.1 which, as discussed later, is typical for luminous
accretion.

It is straightforward to show that the growth of the central black hole by
accretion can have a profound effect on its host galaxy. If the
velocity dispersion of the galaxy is $\sigma$ then the binding energy
of the galaxy bulge, mass $M_{\rm gal},$ is $E_{\rm
  gal}\approx M_{\rm gal} \sigma^2$. The mass of the black hole is
on average observed to be $ M_{\rm BH}\approx 2\times 10^{-3} M_{\rm
  gal}$. If the radiative efficiency of the
accretion process of 10\%, then the energy released by the growth of
the black hole is given by $E_{\rm BH}= 0.1 M_{\rm BH} c^2$. Therefore
$E_{\rm BH}/E_{\rm gal}\approx 2\times 10^{-4}(c/\sigma)^2.$ Very
few galaxies have $\sigma<350\kmps,$ so $E_{\rm BH}/E_{\rm gal} >
100$. The energy produced by the growth of the black hole therefore
exceeds the binding energy of the host galaxy by about two orders of
magnitude!

If even a small fraction
of the accretion energy can be transferred to the gas, then an AGN can have a
profound effect on the evolution of its host galaxy. In practice,
radiation from the accreting black hole will not have any significant
influence on existing stars in the galaxy, It can however strongly
influence gas clouds from which new stars can potentially form. By
ejecting those clouds from the galaxy, AGN feedback can effectively stop
further stellar growth of a galaxy.

The original mechanism of Silk \& Rees is actually a little too
effective and although it predicts a slope to the relation which is
acceptable, the normalization is too low. Gas is blasted out too
easily. However, energy is likely to be radiated away in the process
which is likely to be controlled by conservation of
momentum~\cite{fabian99,2005ApJ...635L.121K,2005ApJ...618..569M}. (Rocket science would be much easier if only energy
conservation were at stake.) It has since been shown that this leads to the
correct normalization and a slightly flatter slope. The details now
centre~\cite{fabian12} on the precise mechanisms responsible for the ejection
(e.g. radiation pressure or accretion disc winds). Energetic winds are
seen from some AGN and UV radiation interacts strongly with dusty gas
in all quasars exerting an outward force through radiation pressure.

It is interesting to note that the effective Eddington limit for dusty
gas exposed to the ultraviolet radiation of a quasar is about 500
times that of ionized gas such as expected close to the quasar. The
$M-\sigma$ relation means that when the quasar is at its Eddington
limit locally, the host galaxy is also at its effective Eddington
limit globally.

Dramatic evidence of AGN feedback occurs in the cores of many clusters
of galaxies. The intergalactic medium in clusters has been squeezed
and heated by gravity to create an intracluster medium which is 10 million K or
hotter. The baryons in the hot gas exceed all baryons in the galaxy
members of the cluster by a factor of more than ten. The most massive
galaxies known lie at the dense centres of clusters where accretion onto their
massive black holes feeds energy into the intracluster medium,
preventing it from cooling down due to the emission of X-rays. The
total mass of gas in a cluster and the depth of the gravitational well
is so large that the gas cannot be expelled. Energy instead flows from
the black hole accretion process into the gas to maintain a
themodynamic status quo.

\begin{figure}[ht]
     \centering
     \includegraphics[width=.8\textwidth]{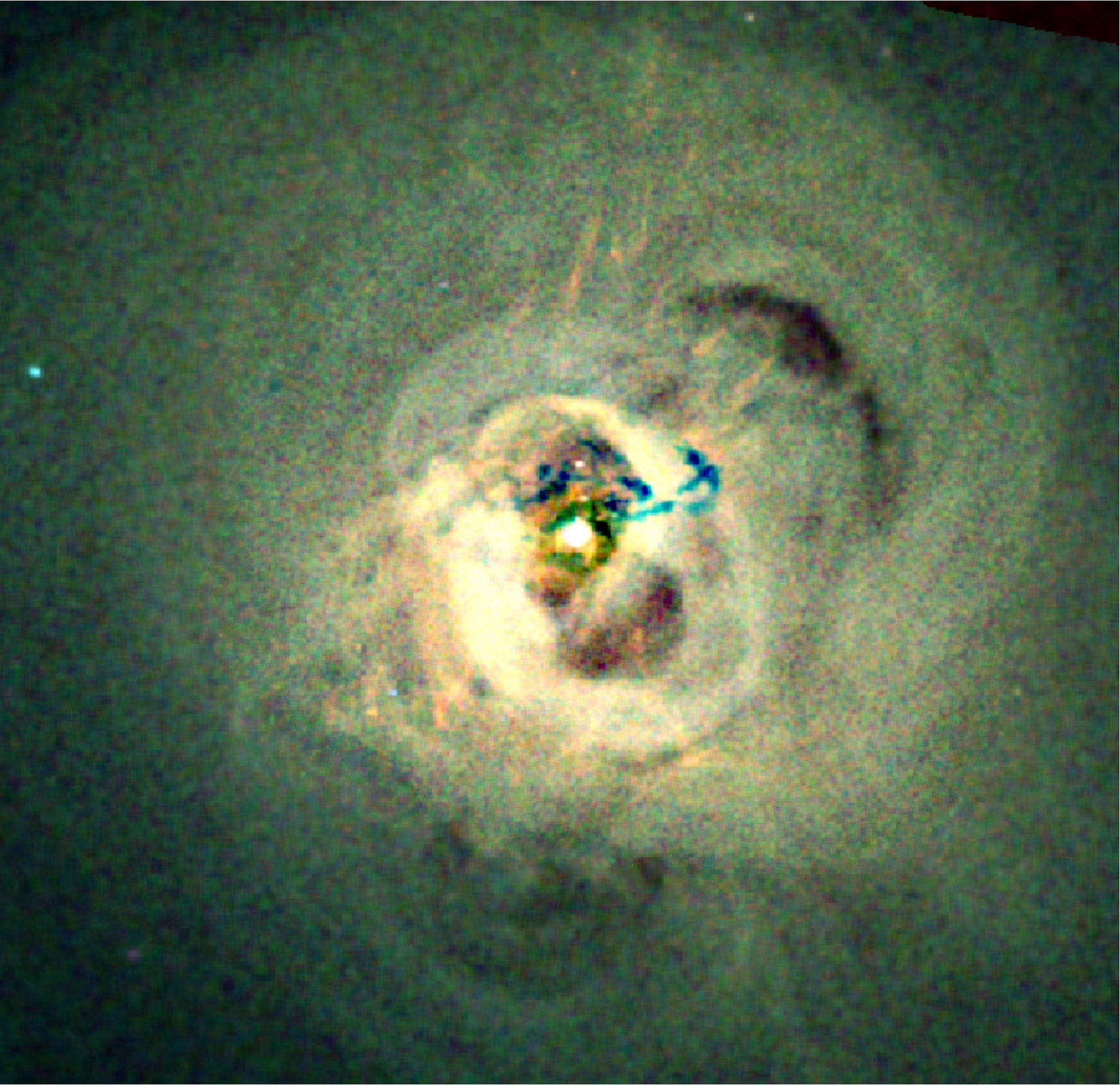}
     \caption{X-ray image of the core of the Perseus cluster centred on the giant
       galaxy NGC1275. The image is about 100 kpc or 300 thousand
       light years across. The AGN at the centre has blown 10 kpc
       diameter bubbles of relativistic plasma into the surrounding
       cluster gas. Detached buoyant bubbles are seen to the NW and S.
       Concentric ripples are seen in the hot gas due to the bubbling
       activity created by the AGN. It is clear that a black hole can
       have dramatic effects on gas structures both in and beyond the
       host galaxy.  }\label{fig:perseus}
\end{figure}

The
energy flows from the near vicinity of the black hole in the form of
relativistic jets which push the hot gas back, forming giant
galaxy-sized bubbles
of relativistic plasma (cosmic rays and magnetic field) --- an example is shown in Fig.~\ref{fig:perseus}.

The bubbles
are buoyant and when large detach and float upward in the cluster
potential, while new bubbles form and grow. The power in the bubbling
process can be estimated from the energy content of a bubble ($4PV$
where $P$ is the pressure of the surrounding gas and $V$ is the volume
of the bubble, the factor 4 being appropriate for relativistic plasma)
divided by the bubble growth time. This power matches the radiative energy loss
of the inner cluster core, indicating that close feedback has been
maintained for billions of years~\cite{churazov02}.

AGN feedback  implies that black holes play a central role in the final
growth and evolution of all massive galaxies. If black holes did not exist then
massive galaxies would be much larger and brighter than we now see.

\subsection{Jets, Gamma-Ray Bursts and the Birth of Black Holes}

\label{sect:jets}

Relativistic jets occur in about 10\% of AGN and in low-state BHB. They
are thus a marker of the presence of a black hole. The details of how
the material is accelerated and collimated are still unclear, with possibilities
discussed later in this Chapter. Apparent superluminal motion
indicates that the bulk Lorentz factor in many jets $\Gamma \sim
10$. Polarization indicates that the radio and much other emission are
due to synchrotron radiation. The jets therefore contain electrons and
a component with positive charge is expected, unless the jets
represent enormous currents. Whether the positive particles are
positrons or protons is not known. If they are protons (or even
heavier nuclei), then the power in the jets can be huge,
exceeding the radiative power in some objects.

Jets are presumably accelerated magnetically by the accretion
disc. Fields emerging from the disc are continuously wound up around
and along the central axis. The details are not yet fully
understood. Jets are seen from all classes of object exhibiting
accretion discs, including young stars and accreting white dwarfs. A
long discussed possibility from quasar jets is that in some cases the
spin energy of the black hole is being tapped by the Blandford-Znajek
mechanism, which is a variant of the Penrose process (see Sections \ref{sect:therm-and-penrose} and \ref{sect:superrad} below). There has been
recent progress in successfully modelling jets using numerical
simulations of this mechanism.

The most relativistic jets occur in Gamma-Ray Bursts (GRB). These were
discovered in the late 1960s by the Vela spacecraft which had a
gamma-ray detector to monitor the Cold War Test Ban Treaty (but not
made public  by Ray Klebesadel and colleagues from Los Alamos until
1973~\cite{klebesadel73}). They are intense flashes of gamma-rays which have a flux which
would make them readily visible to the naked eye if in the optical
band.  Many theories of their origin were proposed, mostly based on
neutron stars. It was not until the late 1990s that a GRB was localized and
identified with a distant galaxy, by the BeppoSax satellite, that made
it clear that GRB are the most luminous events in the Universe,
representing about a Solar rest mass of energy radiated within a few
10s of seconds.

Luminosity and causality point to enormous concentrations of
gamma-rays in a very small region which would instantly lead to
electron-positron pair creation creating a electron scattering dense fog,
preventing any radiation emerging. It is difficult to see how
gamma-rays can be seen unless strong relativistic beaming, $\Gamma \sim
100$, is involved. Many aspects of GRB, including the radio through
X-ray afterglow all support this highly relativistic
interpretation~\cite{piran04}. The source of GRB must also be clean of any other
matter, which could prevent the radiation emerging. It may then be
surprising then that the source of GRB is now believed to be the very
centre of dying massive stars as a black hole is formed. The swirling
matter powers intense jets which burrow through and out of the star,
to produce the GRB display. The birth of a black hole is therefore
marked by one of the most amazing displays in the Universe.

Stars which give birth to black holes are estimated to start life at
masses exceeding $25\Msun$. Such stars have luminosities which are
a million times greater than the Sun and live for just a few million
years.

\section{Current Observations of Accreting Black Holes}

Apart from the nearest examples, we know of the presence of most black
holes from the power emitted as radiation and jets produced as a
result of accretion. Matter falling into a black hole is accelerated
by gravity up to the speed of light and if collisions occur then the
matter can become very hot and radiate. The efficiency of the process
depends on the rate of collisions and on whether the matter has time
to radiate before falling into the event horizon. If the flow is a
thin dense accretion disc then collisions are plentiful and the
efficiency depends on the ISCO (innermost stable circular orbit). Defining the radiative efficiency $\eta$
through $L=\eta \dot M c^2$, then $\eta$ ranges from 5.7\% to 42\% as the spin $a/M$
increases from 0 to 1. In practice it is impossible to spin a black
hole up to exactly 1, and high spins mean $\eta \sim 20-30\%$. Note that $\eta$
for complete nuclear fusion (hydrogen to iron)  is 0.7\%, so black
hole accretion is about an order of magnitude or more higher in efficiency. We now give some details of where these numbers come from, by looking at the orbits of massive particles around a Kerr black hole.

\bigskip

\hrule

{\it

\subsection{Particle motion in the Kerr metric}

\label{sect:kerr-part-motion}

Before discussing the details of particle motion, we need to consider the general structure of the Kerr solution, since this determines the character of the regions in which the particles are able to move.

The most important regions for astrophysical purposes are shown in Fig.~\ref{fig:kerr-structure}.
\begin{figure}[ht]
\begin{center}
\includegraphics[width=0.9\textwidth]{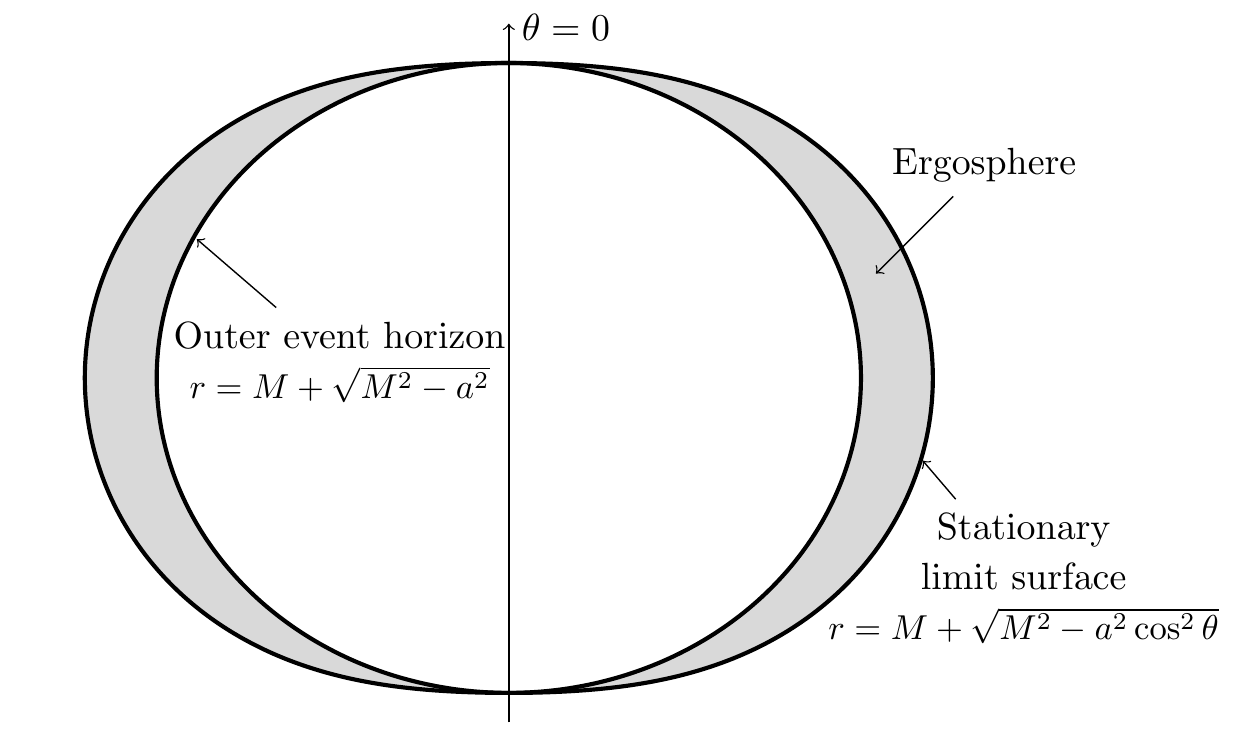}
\caption{The stationary limit and outer horizon surfaces for the Kerr solution.}
\label{fig:kerr-structure}
\end{center}
\end{figure}
We have drawn an outer surface, called the `Stationary limit surface', and the one lying inside this is called the `Outer event horizon'. Both of these are complemented by further surfaces (not drawn) lying closer to the black hole centre (the `Inner Stationary limit' and `Inner horizon' surfaces), but these are shielded by the outer horizon, and so do not have an immediate astrophysical role.

The stationary limit surface marks the point where it is no longer possible for an observer to remain stationary in the $(r,\theta,\phi)$ coordinate system. Try as he or she might, it is inevitable that they will be swept around by the rotation of the hole. Mathematically, this corresponds to the point where the $g_t$ Killing vector corresponding to invariance of the metric under time translations, changes from timelike to spacelike, i.e.\ to where the $g_{tt}$ component of the metric (\ref{eqn:kerr-met}) passes through zero. This requires (setting $c=G=1$ for clarity in the expressions)
\be
r^2-2Mr+a^2\cos^2\theta=0, \quad \text{i.e.} \quad r=M\pm\sqrt{M^2-a^2\cos^2\theta}
\ee
and the outer of these surfaces is shown.

The other surface drawn corresponds to the point where the $g_{rr}$ component of the metric (\ref{eqn:kerr-met}) becomes infinite, and marks the position of the event horizon. This is the position from beyond which a particle can never escape, whatever its mass or state of motion. It is different from the stationary limit, since particles in the inbetween region, although they cannot remain stationary, can in principle escape to infinity. This inbetween region is known as the `ergosphere', since it is possible to arrange particle motions so that work can be extracted from it (see Section~\ref{sect:therm-and-penrose} below). The position of the horizon itself can be found by requiring
\be
\Delta = r^2 - 2Mr + a^2=0, \quad \text{i.e.} \quad r=M\pm\sqrt{M^2-a^2}
\ee
and again the outer of these two surfaces is shown. Given that this radius, unlike the stationary limit surface, does not depend on $\theta$, it might be wondered why the surface is drawn oblate in the diagram. This is because that Boyer-Lindquist coordinates do not correspond to spherical polars even in the case where the mass vanishes, and an embedding of the 2d surfaces we are working with into Euclidean space (for the purposes of visualisation) results in ellipsoids even for $r={\rm const.}$ --- see Section~13.6 of \cite{2006gere.book.....H} for details.

The Lagrangian methods used above in the Schwarzschild case also work here, and one finds quickly the following results for the $\dot{t}$ and $\dot{\phi}$ of a particle moving in the equatorial plane in terms of its conserved specific energy and specific angular momentum, $k$ and $h$:
\be
\begin{aligned}
\dot{t}&=\frac{k\left((r+2M)a^2+r^3\right)-2haM}{r\Delta}\\
\dot{\phi}&=\frac{h\left(r-2M\right)+2 k a  M}{r\Delta}\\
\end{aligned}
\ee

As before, for the $\dot{r}$ equation it is simpler to use the fact that evaluated numerically the interval function $G(x^{\mu},\dot{x}^{\mu})$ is 1 for a massive particle, and 0 for a photon. We start with discussing a massive particle, since our initial emphasis is on the energy liberated from accretion disc orbits. Here we find
\be
\dot{r}^2=k^2-1+\frac{2M}{r}-\frac{1}{r^2}\left(h^2-a^2(k^2-1)\right)
+\frac{2M}{r^3}(h-ak)^2
\ee

We now have three equations we can integrate to find the motion. A key quantity for the energy and stability is the effective potential in $r$. We define this similarly to before via
\be
\frac{1}{2}\dot{r}^2 + U_{\rm kerr}(r)= {\rm const.}
\ee
resulting in
\be
U_{\rm kerr}(r)=-\frac{M}{r}+\frac{1}{2r^2}\left(h^2-a^2(k^2-1)\right)
-\frac{M}{r^3}(h-ak)^2,
\label{eqn:U-func-kerr}
\ee
where an irrelevant overall constant has been ignored.

If we compare with the equivalent in the Schwarzschild case, equation (\ref{eqn:s-V-eff}), we can see that despite the increased complexity of the Kerr metric, the effective potential has terms in just $1/r$, $1/r^2$ and $1/r^3$ as before, resulting in overall similar behaviour (at least for small $a$). A difference, however, is that now the coefficients of the second and third terms depend on the particle energy as well as the angular momentum.

For a circular orbit, we can use the fact that $\dot{r}$ and $\ddot{r}$ are zero to find expressions for the specific energy and angular momentum. For the prograde orbits we are currently considering, these turn out to be
\begin{equation}
k = \frac{1-\frac{2M}{r}+
a\sqrt{\frac{M}{r^3}}}{\sqrt{1-\frac{3M}{r}+ 2a\sqrt{\frac{M}{r^3}}}},
\quad
h = \frac{\sqrt{Mr}-\frac{2aM}{r}
+a^2\sqrt{\frac{M}{r^3}}}{\sqrt{1-\frac{3M}{r}+ 2a\sqrt{\frac{M}{r^3}}}}
\label{eqn:kerr-energy-ang-mom}
\end{equation}

An important quantity for us to find is the radius of the innermost stable circular orbit, expected to be the inner edge of an accretion disc around a Kerr black hole, and the value the energy function takes there.

As in the Schwarzschild case, we can use the effective potential to do this, asking that its second derivative is positive where the first derivative vanishes. Calculating this for the $U(r)$ in (\ref{eqn:U-func-kerr}) leads to the relatively simple criterion for stability
\be
r^2-6Mr+8a\sqrt{Mr}-3a^2>0
\label{eqn:ISCO-r}
\ee
This is soluble analytically for $r$, but perhaps more illuminating to is to look at a plot of $r_{\rm ISCO}$ (the $r$ corresponding to the innermost stable circular orbit) versus spin, as shown in
Fig.~\ref{fig:risco-with-spin}.
\begin{figure}[ht]
\begin{center}
\includegraphics[width=0.8\textwidth]{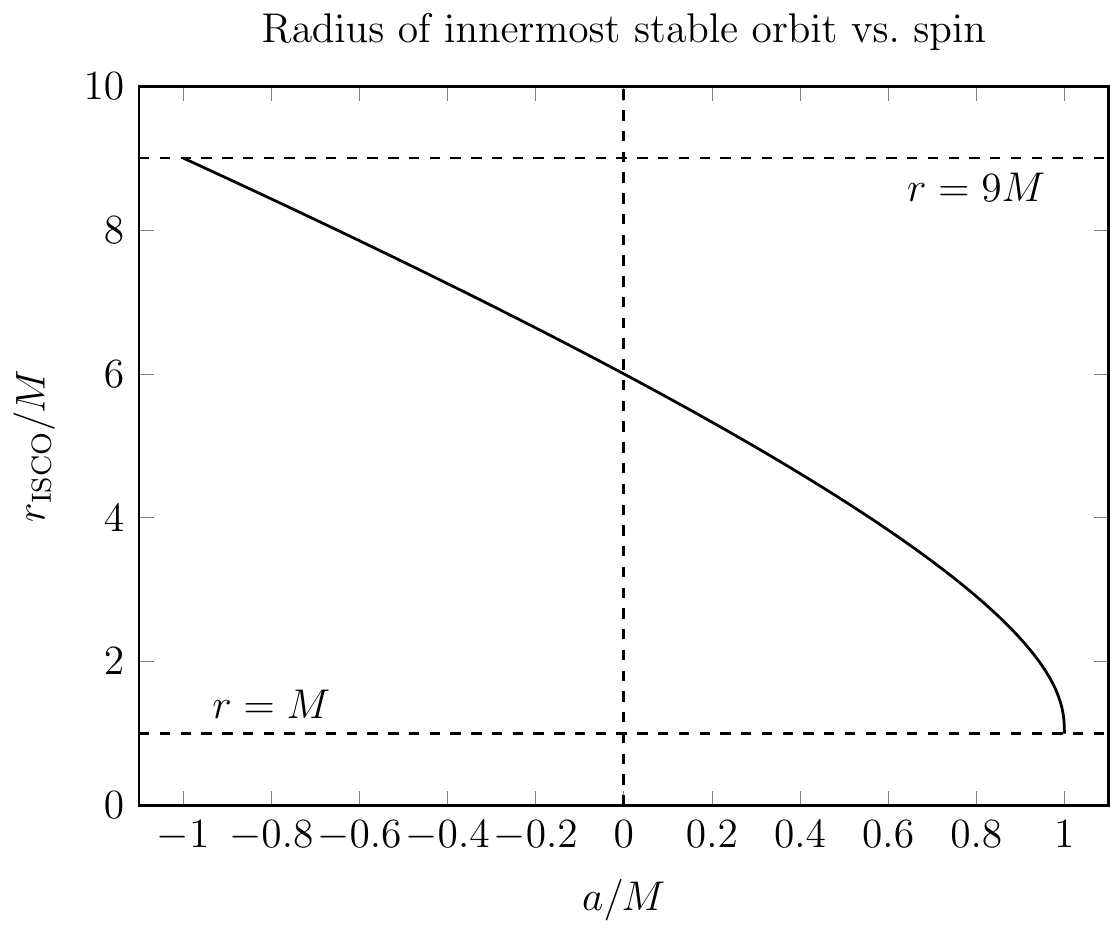}
\caption{Plot of radius of innermost stable orbit vs.\ spin.}
\label{fig:risco-with-spin}
\end{center}
\end{figure}
If $a=M$, an `extremal' black hole, then the solutions are $r=M$ and $r=9M$. If $a=0$, we get the single solution $r=6M$ corresponding to the Schwarzschild case. The former of these shows that stable circular orbits (in the prograde direction) persist right up to the event horizon for a black hole spinning at the maximum mathematically allowed rate, i.e.\ $a=M$. We note, however, that various effects are likely to intervene before this maximum spin rate can be achieved (mainly the counteracting torque felt by the hole in absorbing radiation from the accretion disc \cite{1974ApJ...191..507T}), and $a=0.998 M$ is considered the most likely maximum attainable value, leading to a minimum attainable $r_{\rm ISCO}$ of $1.24 M$.

More generally, by eliminating $r$ between (\ref{eqn:ISCO-r}) and the expression for $k$ in (\ref{eqn:kerr-energy-ang-mom}) we can get a useful relation between the black hole spin and the efficiency of energy release for a particle which has reached the ISCO. For prograde orbits this relation is
\be
\frac{a}{M} = \frac{2\left(2\sqrt{2}\sqrt{1-k^2}-k\right)}{3\sqrt{3}(1-k^2)}
\ee

This provides the plot of efficiency versus spin shown in Fig.~\ref{fig:eff-plot-with-spin}.
\begin{figure}[ht]
\begin{center}
\includegraphics[width=0.8\textwidth]{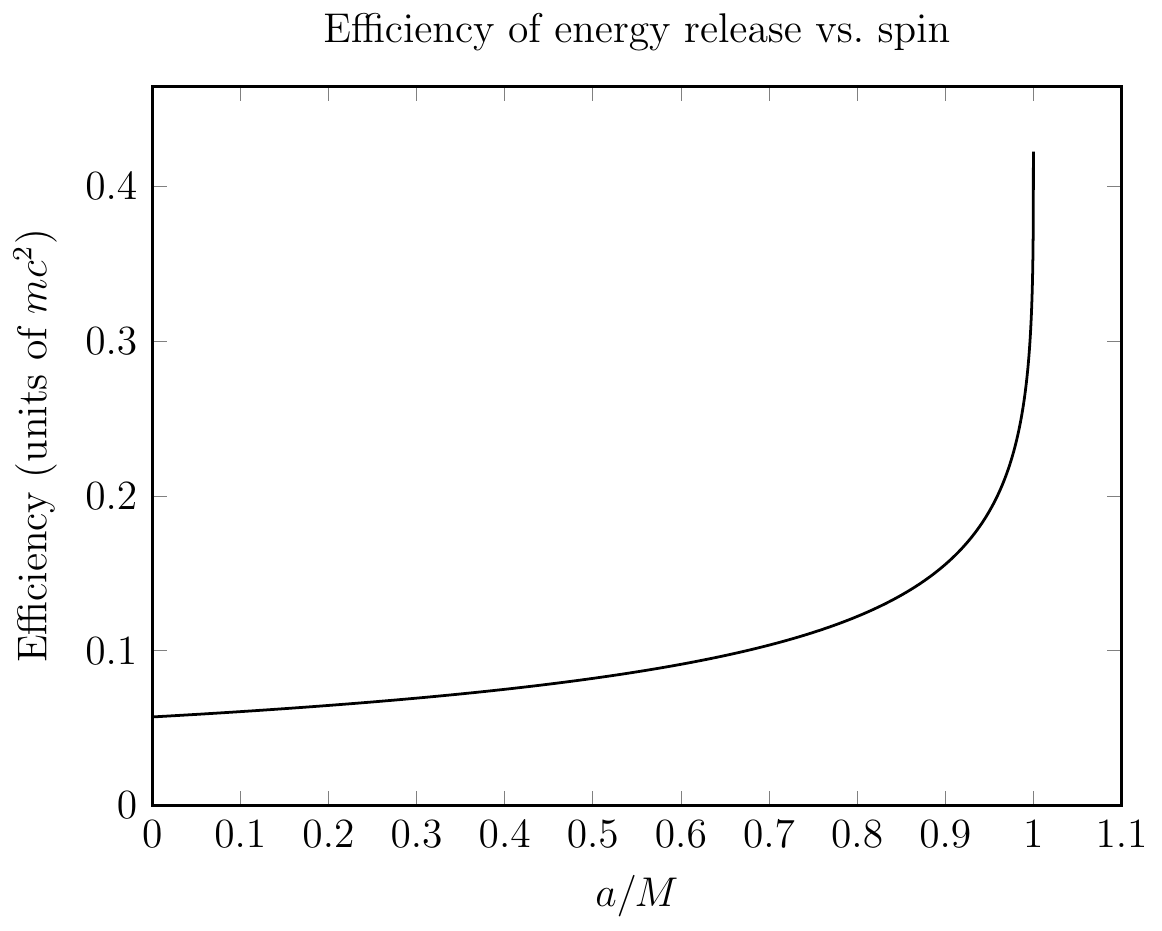}
\caption{Plot of efficiency of energy release vs.\ spin (prograde orbits).}
\label{fig:eff-plot-with-spin}
\end{center}
\end{figure}
Again taking the maximum attainable value of $a/M$ as 0.998, this leads to a maximum attainable efficiency of 32\%.

}

\bigskip

\hrule

\subsection{Current Observations of Accreting Black Holes contd.}

The accreted matter can be supplied by a companion star, in the case
of a BHB, or from surrounding gas in AGN. How much is captured by the
black hole in the AGN case will depend on the density, temperature,
magnetic field and angular momentum of the gas. The case when there is
no angular momentum and magnetic fields can be ignored is known as
Bondi accretion, which assumes that the gas can be treated as a
fluid. Basically gas within the Bondi radius, defined as the radius
beyond which the gas particles have the escape velocity, will be
captured and this can be many orders of magnitude larger then would
occur if the particles were free particles with no interactions or
collisions.  Zero angular momentum is a rather extreme assumption and
may not apply to any real system, but it presents an easily calculated
order of magnitude estimate of the accretion rate. Angular momentum is
likely to reduce any inflow rate and could even choke off the
flow. Indeed accretion flows can be unstable when heating by radiation
from the central accreting mass is considered. Magnetic fields and
turbulence also contribute to complexity and accretion flows are
commonly observed to show chaotic variability.

The accretion disc will be thin, with the ratio of thickness to
radius, $h/r<0.1$, when the luminosity is above
about 1\% $L_{\rm Edd}$. At lower accretion rates the disc thickens as the
collision rates drop which can then cause the efficiency to drop with
further decreases in $\dot M$. (The black hole swallows both the matter and energy released.) The properties of such radiatively
inefficient flows (RIAFS)
depend on whether the electrons (which radiate) and the protons (which
have much of the energy) remain coupled, i.e.\ have the same
temperature. There are still theoretical issues remaining as to whether
this happens in such a highly magnetized plasma, but observations do
indicate changes in behaviour in objects with low accretion rates. Not
least is a tendency for jets to become common. The low luminosity of Sgr A* suggests it is in an extreme RIAF state.

The luminous accretion flows, however, are usually above 1\% $L_{\rm Edd}$ and so
are thin and dense and optically thick resulting in  much radiation
being emitted as
a black body.  The spectrum of the disc is then produced by the sum of
blackbodies from different radii which gives a $\nu^{-1/3}$ spectrum at
low energies up to the highest energies, which come from the smallest
radii, where it turns over into a Wien tail. The peak temperature
scales as $\dot M r_{g}^{-{1/4}}$. This means that the most massive luminous black
hole systems have spectral peaks in the UV, while stellar mass black holes
peak in the X-ray band.

Significant power can also emerge as a results of magnetic fields
wound up in the accretion flow. They can emerge from the disc, become
twisted and create a mess of reconnecting field lines powering a hot
corona, with a temperature which can easily be 100 times hotter then
the disc itself.  The million degree Solar corona with associated flares
and outflows is a nearby example but even low luminosity AGN have a
power $10^{12}$ times larger than the Sun, so the analogy should not
be taken too
literally. The corona is the source of most of the X-rays seen from
AGN.  Evidence discussed later indicates that the corona in bright
objects is located above and close to the black hole ($\sim 10 r_{\rm g}$). The
magnetic field structure also appears to power jets in some objects,
and outflows in others. The degree to which the spin of the black hole
is important (and maybe also a small ISCO) are still debated.

BHB have stellar-mass black holes and accrete from their companion
star. Disc formation is implicit due to the orbital angular momentum.
A few  sources such as Cyg X-1 are persistent, with their
long-term luminosity varying by less than an order magnitude, but
sometimes switching states from hard (coronal dominated) to soft
(thermal disc dominated). Many of the BHB are however transient,
having outbursts every few years or decades, during which the
luminosity can rise from $10^{33}$ to $10^{39}\ergps$. Outbursts follow a
similar pattern in the luminosity-colour (hardness) plane, starting
hard as the luminosity builds up towards $L_{\rm Edd}$, then switching to being
soft as the thermal disc emission dominates over the corona, before
subsiding back to the hard state as they decline to quiesence~\cite{remillard06}.  At the
peak of an outburst the nearest of these sources (eg A0620-00 at 2.5 kpc)
can become the brightest in the sky, outshining Sco X-1 and even
the Sun (except when it is flaring) by a large factor.

There are also many similar X-ray binaries in our Galaxy which host a
neutron star instead of a black hole. We know about the neutron star
by observation of rapid pulsations due to the spin of the highly
magnetic field, or from X-ray bursts which are thermonuclear flashes
taking place very near the neutron star surface. Many of these are
also transient and behave in a similar way to BHB. The depth of
quiesence has been proposed as evidence of an event horizon in BHB by
Ramesh Narayan and colleagues~\cite{narayan08}. Black holes have no hard surface and
can accrete matter and radiation whereas neutron stars must emit the
gravitational energy released. BHB generally have lower quiesent
luminosities than neutron star binaries which fits in with this
picture. Whether these systems have similar enough accretion rates in
quiescence for this to be a proof has yet to be determined.

Both BHB and some neutron star systems also show quasi-periodic
oscillations, discovered by Michiel van der Klis in 1986~\cite{vanderklis00}. This
phenomenon is complex and may be a rich source of information about the
accretion flow and geometry, but it has yet to be fully understood and
interpreted. To understand at least one of the frequencies we would expect to be seen, we now calculate the expected orbital frequency seen by a distant observer for material in the innermost orbit.

\bigskip

\hrule

{\it

\subsection{Velocities and frequencies}

For astrophysical black holes, further important quantities for circular orbits are the velocity in the orbit, and the orbital period. For the former, the velocity determines (essentially via special relativistic effects) an important component of the shift in frequency which occurs for line radiation coming from the accretion disc, and for the latter, we can in some cases see directly the modulation of radiation associated with an orbital period. These are thought to provide one of the frequencies within the `quasi-periodic oscillations' which are occasionally observed from accreting stellar mass black holes and can have frequencies in the several hundred Hz region \cite{vanderklis00}.

Working with a Schwarzschild black hole, then if we let $v$ be the `ordinary' velocity in a circular orbit, and $\alpha$ be the corresponding `rapidity', then (temporarily dropping factors of $c$ since this makes the formulae clearer), we have
\be
v=\tanh\alpha \quad \text{and} \quad r\dot{\phi} = \gamma v = \sinh\alpha
\ee
We can thus use (\ref{eqn:eqn-for-h}) for $h$ and the definition $r^2 \dot{\phi} = h$, to calculate $v$ as a function of $r$, obtaining the simple result
\be
v^2 = \frac{GM}{r-2GM}
\ee
For an object rotating in the innermost stable circular orbit (ISCO), at $r=6GM$, we therefore find that its velocity is half the speed of light. (Note that, perhaps surprisingly, exactly the same velocity is obtained for the ISCO about a spinning black hole with any magnitude for its spin.)

As regards orbital frequency, then since the $t$ coordinate is the time as measured by a stationary observer at spatial infinity, if we can form $d\phi/dt$ then setting this equal to $2\pi \nu_{\rm orb}$ will immediately give us the orbital frequency, in Hz, as measured by an external observer.

Thus using
\be
\frac{d\phi}{dt} = \frac{d\phi}{d\tau}\frac{d\tau}{dt} = \frac{\dot{\phi}}{\dot{t}},\quad
\dot{\phi} = \frac{h}{r^2}=\sqrt{\frac{GM}{r^3}\frac{1}{(1-\frac{3GM}{r})}}
\ee
and
\be
\dot{t} = \frac{k}{1-\frac{2GM}{r}} = \frac{1}{\sqrt{1-\frac{3GM}{r}}},
\ee
we find
\[
\frac{d\phi}{dt}=2\pi \nu_{\rm orb}=\frac{\dot{\phi}}{\dot{t}}=\sqrt{\frac{GM}{r^3}},
\]
which despite being derived in full GR is exactly the Newtonian (Keplerian) result.

As an example, evaluating at the innermost stable circular orbit, $r=6GM/c^2$, for a 10 solar mass black hole, yields the impressively high result of $\nu_{\rm orb}=218 {\rm \, Hz}$. Thus an emitting blob orbiting at this radius could potentially produce this frequency in the frequency spectrum of the source.

}

\bigskip

\hrule

\subsection{Further AGN properties}

AGN range in luminosity from below $10^{10}$ to $10^{14}\Lsun$ and are
situated at the centre of the host galaxy. They appear to respect the
Eddington limit and show strong evolution in luminosity with redshift
(and thus cosmic time). The quasar and therefore mass-building phase
peaked at about redshift 2.5 with quasars being observed back to
redshift 7.1. An important issue with AGN is that a plentiful supply
of fuel leads to obscuration of the source itself. If the absorbing
column density is Thomson thick (ie most photons are scattered by
electrons on passing out of the source), then only heavily reprocessed
emission is all that remains, emerging predominantly in the infrared.

Type I AGN are largely unobscured and Type II obscured. The difference
may be geometrical and due to our line of sight. In a broad class of
source the obscuring medium is in the form of a thick torus at a
radius of about a parsec. If viewed in the plane of the torus we see a
Type II obscured object and if we view it face-on then it is Type
I. If at the same time there is a jet, then face-on down the jet the
object appears highly beamed as a blazar~\cite{urry95}.

\section{Measurements of the Masses of Black Holes}

The mass of a compact object can be measured from the motions of stars
or gas in its vicinity using Newton's Law of Gravitation. If a star
makes a circular orbit around a massive black hole with velocity $v$ at
radius $r$ then the mass of the black hole $M=v^2 r/G$. Multiplying
factors emerge for different shaped orbits but the essence of measuring
velocity and radius remains the same. For the black hole at the centre
of our Galaxy, Sgr A*, the 3D motion of many nearby stars has been mapped
to give the best estimate of $4\times 10^6\Msun$.

With stellar mass black holes orbiting a normal star, then constraints
on the mass are obtained by measuring periodic doppler shifts of
spectral lines of that star to yield its radial velocity curve of
semi-amplitude $K$ and period $P$. These are combined to give the mass
function
\be f={{(M_{\rm bh}\sin i)^3}\over{(M_{\rm s}+M_{\rm bh})^2}}={{P K^3}\over{2\pi G}}\ee

Further progress requires estimates of the mass of the
normal star, perhaps from its spectral type, and of the inclination of
the orbit $i$. If the system is eclipsing then $i$ must be high, but
if not then other means must be used such as observation of periodic
ellipsoidal variations in the lightcurve of the star due to tidal
distortions. These can be modelled to yield the inclination.  There
are now several dozen black holes with masses estimated to about 10\%
in this manner.

The mass of supermassive black holes (SMBH) is obtained from the motions of
nearby stars and gas if they lie in relatively nearby galaxies where the
region of influence of the black hole can be resolved. This is roughly
the radius within which the black hole-induced motions are faster than the
general motions in the galaxy. The masses of more distant active SMBH can be
determined using the reverberation technique. Delayed variations in the
ionizing UV radiation from the nucleus are seen in orbiting
broad-line clouds --- the whole system appears to reverberate with delays
ranging from days to weeks. (The
clouds show doppler-broadened emission lines due to orbiting close to the SMBH.)
The light travel-time delay then gives the radius and the line width  of
the clouds gives the  velocity required to determine the central mass. Assumptions need to made about the
geometry but careful work of the past two decades has established the
success of the method~\cite{peterson13}. It has enabled correlations to be built between
the width of selected emission lines, optical luminosity and black
hole mass.

Current results indicate a maximum mass for observed neutron stars at
just over $2\Msun$, while stellar mass black holes have masses from
$\sim 4-20\Msun$, with an average of about $7\Msun$. Observed
supermassive black holes have masses ranging from just below $10^6$
to $2\times 10^{10}\Msun$. It is possible that some UltraLuminous X-ray
sources (ULXs), which have luminosities greater than the Eddington
limit for a $10\Msun$ black hole, are Intermediate Mass Black Holes
(IMBHs) with masses of $10^2-10^5\Msun$.

\section{Measurements of Black Hole Spin}

Whilst mass measurements can be made at a large distance from an
object, spin requires a probe which is close in, within $10 r_{\rm g}$\footnote{If the mass accretion rate can be measured (from
  observations of the outer disc, say) and the total luminosity is
  known, then in principle spin can be deduced from the radiative
  efficiency using the relation in Fig.~\ref{fig:eff-plot-with-spin}.}. The
standard approach is to identify the inner edge of the accretion disc
with the ISCO and then convert that radius to spin. This means that
the inner disc must be luminous and detectable.

Bright Galactic X-ray binaries in the soft state generally have a
quasi-blackbody X-ray spectrum which can be used to measure the
emitting area using the Stefan-Boltzmann law in an analogous way to that
by which stellar radii are determined (blackbody luminosity $L=4\pi R^2\sigma T^4$)~\cite{mcclintock11}.
Unlike the surface of a star, an accretion disc is not all at the same
temperature, so the emission profile of the disc must be assumed and the black hole
mass and distance determined in order to measure the ISCO and thus
spin.

\begin{figure}[ht]
     \centering
     \includegraphics[width=.8\textwidth]{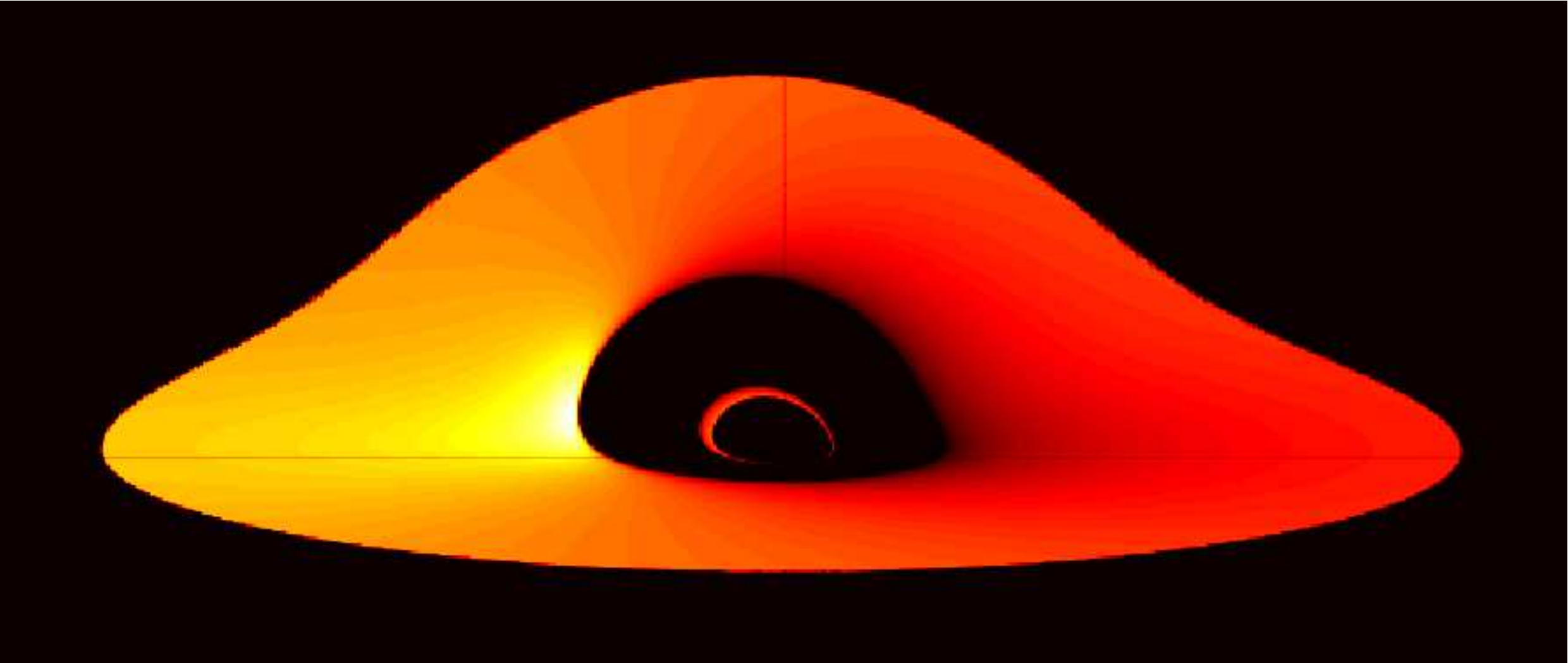}
     \caption{Pseudoimage of a luminous thin accretion disc around a
       black hole with doppler and gravitational redshift effects
       included. Light bending causes the distant side of the disc to
       appear as an arch above the hole. The smaller loop near the centre is due to scattering of photons from the (unstable) photon orbit at $r=3GM/c^2$. The colour scale indicates flux received.}
\label{fig:accretion-disc-image}
\end{figure}

Light bending effects are very significant for the accretion disc near a black hole~\cite{miniutti03}, so we now give some details of light bending effects in GR, first in the Schwarzschild metric, and then for Kerr black holes. An impression of what we would actually see for the accretion disc around a Schwarzschild black hole, if sufficient resolution was available, is shown in Fig.~\ref{fig:accretion-disc-image}.

\begin{figure}[ht]
     \centering
     \includegraphics[width=.8\textwidth]{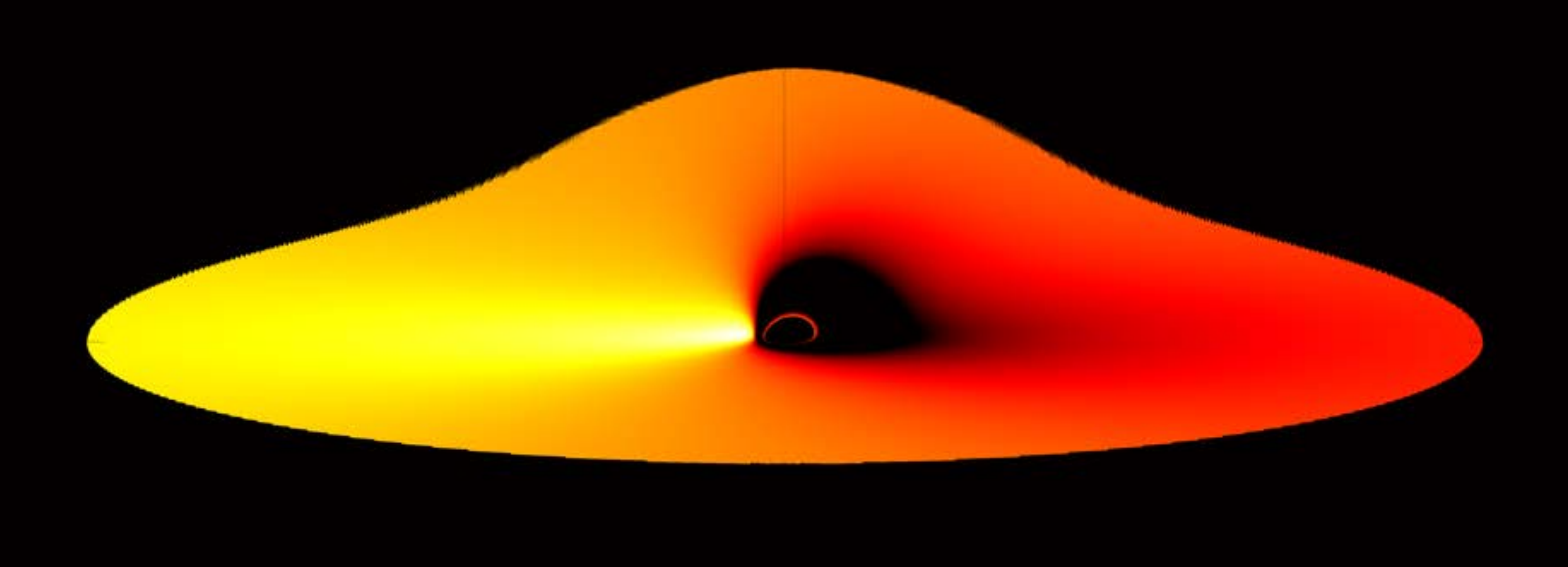}
     \caption{As for Fig.~\ref{fig:accretion-disc-image}, but for a Kerr black hole rotating at the maximum attainable value of $a=0.998 M$ (see Section~\ref{sect:kerr-part-motion}). The unstable photon orbit is now near $r=GM/c^2$.}
\label{fig:accretion-disc-extreme}
\end{figure}

This can be compared with the more distorted and asymmetric image we would see from a black hole rotating at the expected maximum rate of $a=0.998 M$, in Fig.~\ref{fig:accretion-disc-extreme}.

\bigskip

\hrule

{\it

\subsection{Equations for photon motion and redshift}

The other results we need for understanding astrophysical aspect of black holes, are the equations of motion for photons, and results for how their energy changes during propagation.

For the equations of motion, we can use the above analysis, but with the interval function $G(x^{\mu},\dot{x}^{\mu})$ set equal to 0 rather than 1, once its functional variation has been taken.

Tracking through the changes this causes, we find equation (\ref{eq-for-rdot}) is replaced by
\begin{equation}
\dot{r}^2=k^2 c^2-\frac{h^2}{r^2}\left(1-\frac{2GM}{rc^2}\right)
\end{equation}
and for the `shape' equation, (\ref{eq-for-u}), one finds that it is now the `Newtonian' term that disappears on the r.h.s., and we just get
\be \f{d^2u}{d\phi^2} + u = \f{3GM}{c^2}u^2.
\label{eqn:photon-shape}
\end{equation}
One can immediately confirm from this equation that there is a circular photon orbit at $r=3GM/c^2$, as mentioned above, but is it stable?

To do the stability analysis, we rewrite the energy equation as
\be
\frac{\dot{r}^2}{h^2} + V_{\rm eff}(r) = \frac{1}{b^2},
\label{eqn:photen}
\ee
where $b=h/ck$, $\mu=GM/c^2$ and the effective potential
\be
V_{\rm eff}(r) = \frac{1}{r^2}\left(1-\frac{2\mu}{r}\right).
\label{eqn:V-eff}
\ee
Let us look at a plot of this function, Fig.~\ref{fig:phot-V-eff}.
\begin{figure}[ht]
\begin{center}
\includegraphics[width=0.7\textwidth]{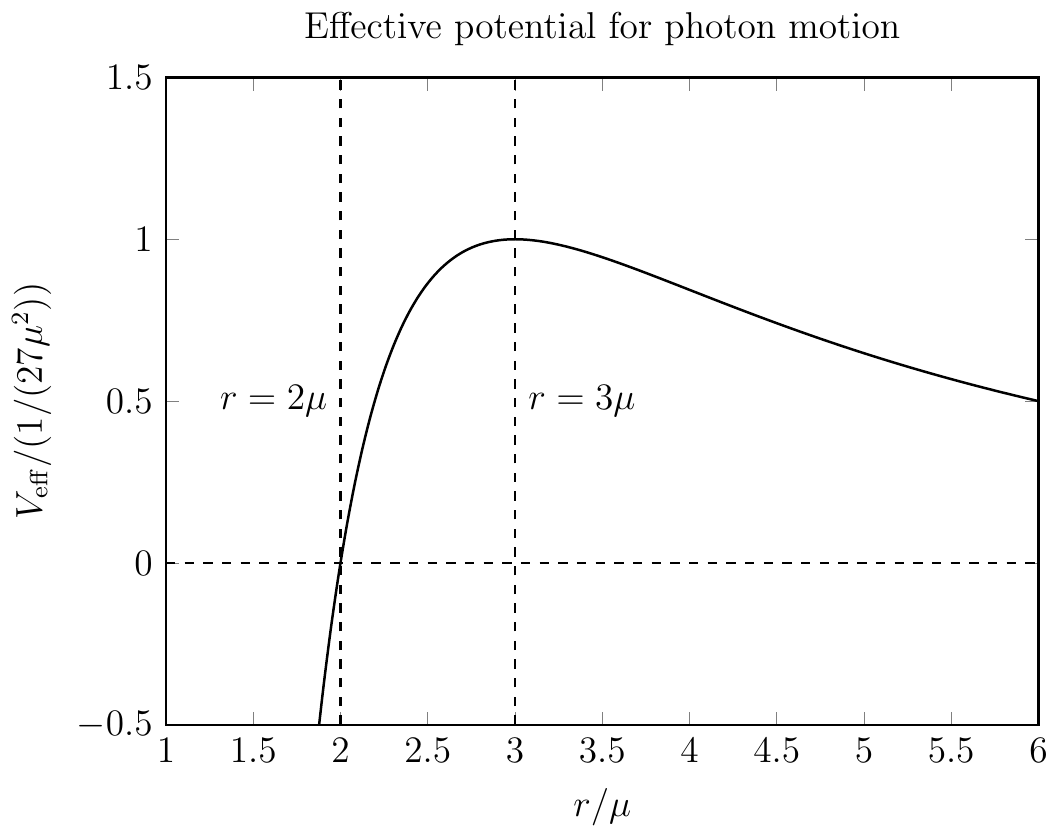}
\caption{Plot of the effective potential for photon motion in the Schwarzschild geometry. (Units of $V_{\rm eff}$ are $1/(27\mu^2)$ where $\mu=GM/c^2$.)}
\label{fig:phot-V-eff}
\end{center}
\end{figure}
We can see $V_{\rm eff}(r)$ has a single maximum at $r=3\mu$ where
the value of the potential is $1/(27\mu^2)$. This shows that the circular orbit at $r=3\mu$ is unstable and in fact no stable circular photon orbit is possible in the
Schwarzschild geometry.

Another immediate use of equation (\ref{eqn:photon-shape}) is in connection with light bending.
Referring to Fig.~\ref{fig:light-deflect}, we can see that a suitable first solution in which the term
$3GMu^2/c^2$ is completely ignored, is
\be
u=\f{\sin\phi}{R},
\label{eqn:u-straight}
\end{equation}
where $R$ is the radius of the body the gravitational deflection due to which
we wish to work out.
\begin{figure}[ht]
\begin{center}
\includegraphics[width=0.7\textwidth]{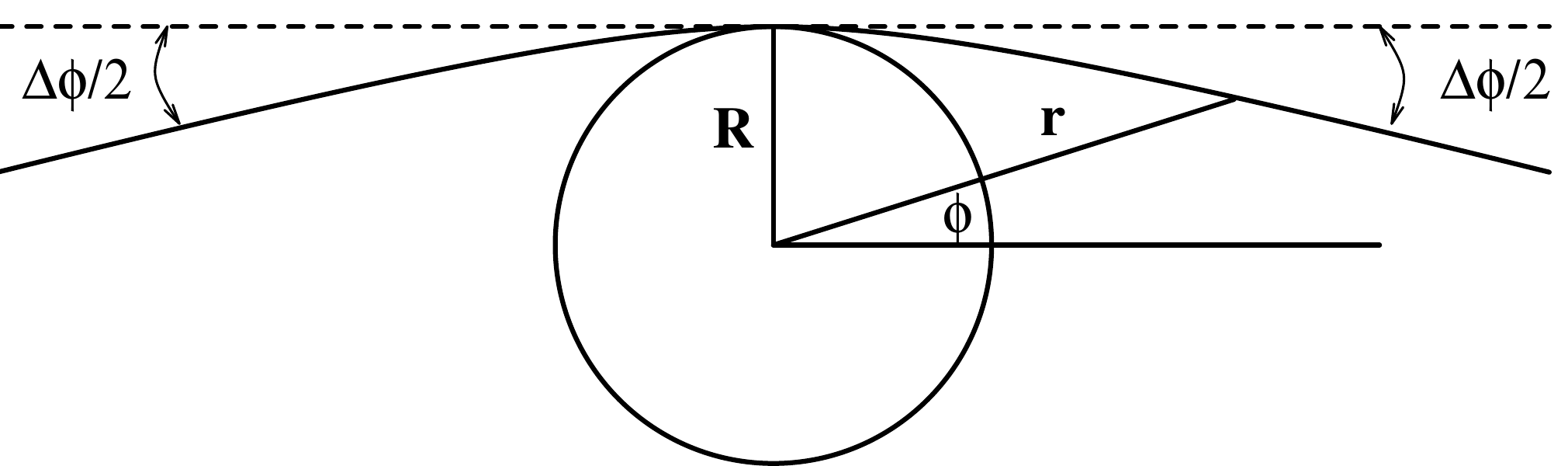}
\caption{\sl Gravitational deflection of light}
\label{fig:light-deflect}
\end{center}
\end{figure}
We iterate this equation by putting $\sin^2\phi/R^2$ for $u^2$ on the r.h.s. of
(\ref{eqn:photon-shape}), i.e.
\be
\f{d^2 u}{d\phi^2} + u = \f{3GM}{c^2R^2}\sin^2\phi.
\end{equation}
This is satisfied by the particular integral
\be
u_1 = \f{3GM}{2c^2R^2}\left(1+\f{1}{3}\cos2\phi\right),
\end{equation}
and adding this into the original solution yields
\be
u = \f{\sin\phi}{R} + \f{3GM}{2c^2R^2}\left(1+\f{1}{3}\cos2\phi\right).
\end{equation}

Now consider the limit $r\rightarrow \infty$, i.e. $u\rightarrow 0$. Clearly we
can take $\sin\phi \approx \phi$, $\cos 2\phi \approx 1$ there, and we obtain
$\phi = -2GM/(c^2R)$ so that the total deflection (see figure) is
\be
\Delta \phi =\f{4GM}{c^2 R}.
\end{equation}
This is the famous gravitational deflection formula. For the Sun it yields 1.77
seconds of arc, and was first verified in the 1919 eclipse expedition. More
recent high precision tests use {\em radio sources}, since these can be
observed near the Sun, even when there is no eclipse, and there is now no doubt
that the GR prediction (which incidentally is twice what had previously been
worked out using a Newtonian approach) is accurate to a fraction of a percent.

In the black hole context, much stronger bending can occur, for which this first approximation is insufficient, and the full equations of motion need to be integrated. This can lead to some interesting effects, illustrated below in the context of a Kerr black hole.

The other important aspect to discuss is gravitational redshift. Quite generally, for a metric with a Killing vector field $g_t$, the redshift due to gravitational time dilation will be given by the ratio of $p \! \cdot \! g_t$ at emission and reception, where $p$ is the photon 4-momentum. $g_t$ becoming null, i.e.\ $g_{tt}$ going to zero, marks the point where infinite redshift is possible since $p$ can then be in the direction of $g_t$, e.g.\ for counter-rotating photons at the Kerr Stationary limit. This redshift from $p \! \cdot \! g_t$ acts multiplicatively on any further redshifts which are occurring at a point due to special relativistic effects. E.g.\ for a point in an accretion disc, there will be a doppler redshift due to its motion relative to the initial path of an emitted photon, and then this will be multiplied by the gravitational redshift the photon experiences between that point and infinity. The combined effects of the redshifts can be seen in pattern of received flux seen in Figs.~\ref{fig:accretion-disc-image} and \ref{fig:accretion-disc-extreme}, which are for the case where the accretion disc itself has constant emissivity.

\subsection{Light bending around a Kerr black hole}

The treatment of photon orbits for the Kerr case can proceed in the same way as in Section~\ref{sect:kerr-part-motion}, which dealt with massive particles, except that now the interval function $G(x^{\mu},\dot{x}^{\mu})$ evaluates to 0 instead of 1. Several textbooks, e.g.\ \cite{2006gere.book.....H} and \cite{frolov2011introduction}, deal with the details of the resulting orbits, and in particular show that as in the Schwarzschild case, there are still no stable circular photon orbits.

The possible orbits for non-equatorial photons can be quite exotic. As an example, we show in Fig.~\ref{fig:kerr-photon-track} the path of a photon emitted from an accretion disc in the equatorial plane, moving towards the black hole but with an initial upward component to its velocity. This moves round behind the black hole and is eventually able to escape to infinity, moving in a roughly opposite direction to its initial motion. This example is for a Kerr black hole with the maximum expected value for $a$ of $0.998 M$ --- the ISCO and horizon radii are also shown.
\begin{figure}[ht]
\begin{center}
\includegraphics[width=0.7\textwidth]{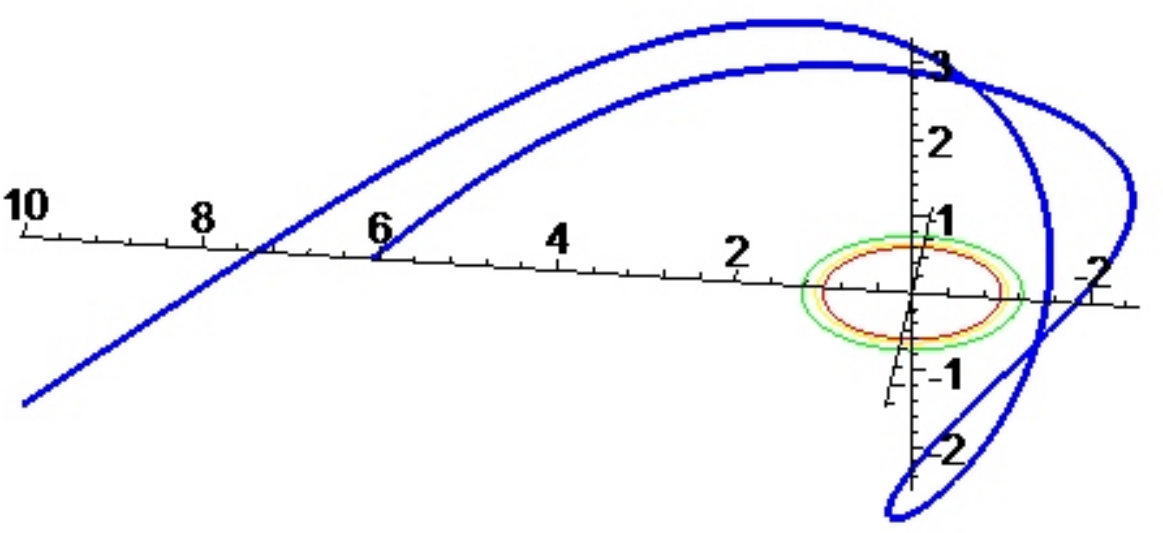}
\caption{\sl Path of a photon emitted from an accretion disk that eventually escapes from the black hole in a direction about 180$^\circ$ from its initial direction.}
\label{fig:kerr-photon-track}
\end{center}
\end{figure}

}

\bigskip

\hrule

\bigskip

\subsection{Iron Line Emission}

\begin{figure}[ht]
     \centering
     \includegraphics[width=.8\textwidth]{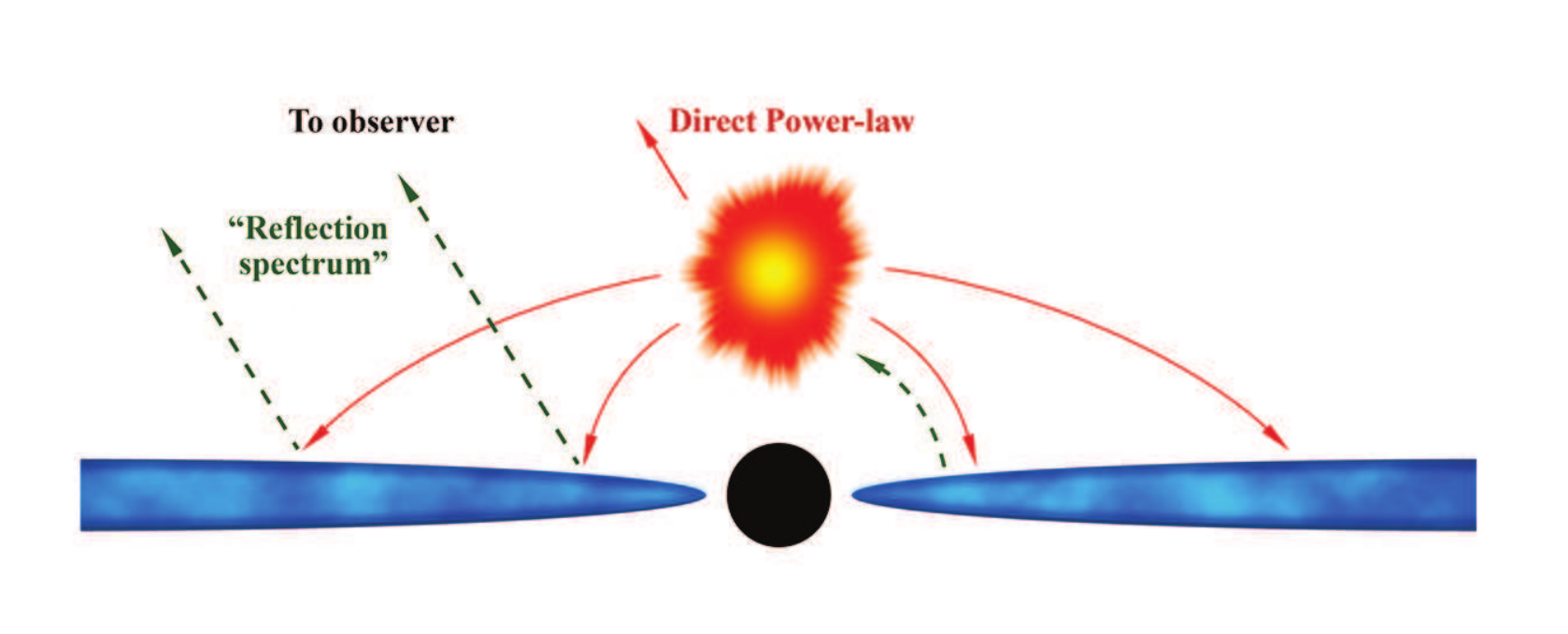}
     \caption{Schematic cross-section of an accretion disc
       around a black hole with a central corona providing
       irradiation. The production of the reflection spectrum is
       illustrated. }
       \label{fig:corona}
\end{figure}

The method discussed above for Galactic X-ray binaries does not work for AGN since the much higher black hole
mass means that the  blackbody disc emission emerges in the
far UV band which is unobservable because of absorption by Galactic hydrogen clouds. Observations are made instead of broad iron-line emission produced by
X-ray reflection. This is the fluorescent and back-scattered emission
produced by irradiation of the disc by the coronal X-ray continuum --- see Fig.~\ref{fig:corona}.
X-rays
absorbed in the disc lead to line emission, particularly iron K
emission at 6.4 to 6.95 keV, depending on the ionization state of the
emitter. At higher energies the X-rays are more likely to be Compton-scattered
back out of the disc rather than absorbed, producing what is called
the Compton hump in the reflection spectrum.

\begin{figure}
     \centering
     \includegraphics[width=.7\textwidth]{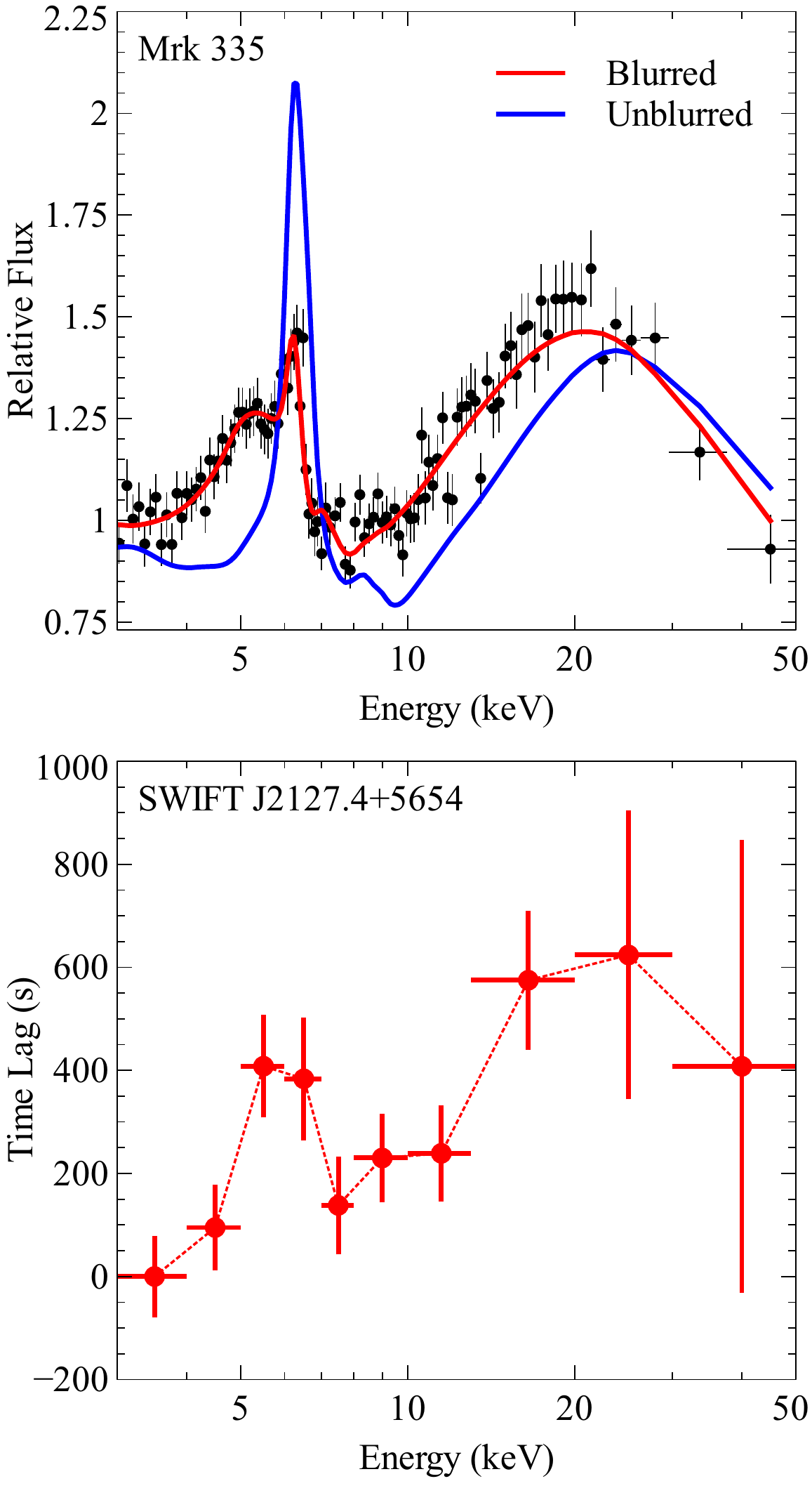}
     \caption{Upper: The unblurred reflection spectrum including the
       iron-K emission line and Compton hump (solid line) appears
       relativistically blurred to the observer as shown in the line
       fitted to the data points. The obvious dominant redshift is
       gravitational due to the proximity to the black hole. The
       source is the Seyfert galaxy Mkn335, observed by NuSTAR during
       a low state. Modelling of the data shows that the black hole in
       Mkn335 has a spin $a/M>0.97$. Most of the emission seen in the
       spectrum emerges from within $2\rg$ of the event horizon \cite{parker14}. Lower:
       Time lags measured by NuSTAR from the AGN SWIFTJ2127 showing both
       iron K and Compton peaks.  }
       \label{fig:nustar}
\end{figure}

The spectral features are produced in the innermost parts of the
accretion disc where gravitational effects are strong and lead to large
doppler shifts and gravitational redshifts which skew and broaden
emission lines~\cite{fabian89,fabian14} --- an example is shown in Fig.~\ref{fig:nustar}.
Detailed modelling of the line shape, and generally
the shape of the whole observed reflection spectrum, leads to an
estimate of the inner radius of the disc and thus the spin. Radii are
determined in this method directly in gravitational units so the
distance and mass of the black hole are not required.

The broad iron line method can also be used in certain states of
BHB~\cite{miller07} and has been found to give consistent results with the blackbody method.

Confirmation that the reflection modelling is correct has recently
been obtained from X-ray reverberation~\cite{uttley14}. This involves the small
(tens of seconds to tens of minutes) time delays or lags whch occur between
the detection of an intrinsic change in luminosity of the corona and
the corresponding change in the reflection. The energy spectrum of the
high frequency time lags shows a broad iron line indicative of
reflection (lower panel of Fig.~\ref{fig:nustar}).

The results for a number of objects imply that high spin
is common for the brightest observed sources and the corona in those
objects is compact and at a height of $5-10r_{\rm g}$. In this regime, strong light
bending occurs such that much of the coronal emission is bent down to
the disc making the reflection strong. Month-long temporary episodes
have been seen in a few sources in which the continuum drops to being
almost unobservable, with only slight changes to the reflected
spectrum. It appears that the corona has dropped below $5r_{\rm g}$ in these
cases and that light bending is preventing much of the continuum being
observed at all. Most of the reflection spectrum is then emerging from
within $2-3r_{\rm g}$, which is the closest emission easily seen from around
any black hole~\cite{parker14}.

Further evidence of a compact corona is obtained from microlensing
studies of distant quasars which are lensed by intervening galaxies.
This produces several images of the quasar which follow different
light paths through the galaxy. Stars in the galaxy crossing one of
those paths amplify the image by lensing to a degree depending on how
the angular sizes of the emission region of the quasar and
the Einstein ring of the star match one another. For typical values this
means that higher amplitude variability is expected from smaller
emission regions. Both optical and X-ray observations have been made
of several lensed quasars which show that the X-ray emission region is
much smaller than the optical one, and in several cases is less than
$10r_{\rm g}$ in size~\cite{blackburne14}.

\section{Future Observations of Astrophysical Black Holes}

\label{sect:future-obs}

Many exciting developments are expected over the next decade or so in
observational studies of black holes. One important step will be the
networking together of several millimetre-wavelength telescopes to
form the Event Horizon Telescope which will resolve the ``shadow'' of
the black hole in Sgr A*~\cite{doeleman09}.  Extreme light bending of the emission from the accretion flow onto the black hole should reveal a dark
patch, the shadow, at the position of the black hole. The diameter of
the shadow is determined by the paths of light rays emitted from just
above the event horizon to the observer and corresponds to about $10\rg$
almost independent of the black hole spin.

\bigskip

\hrule

\bigskip

{\it

{\bf Calculation of the shadow radius:}

We use the results above for the photon effective potential in the Schwarzschild case. Putting together equations (\ref{eqn:photen}) and (\ref{eqn:V-eff}), we find
\be
 \frac{\dot{r}^2}{h^2}+\frac{1}{r^2}\left(1-\frac{2\mu}{r}\right) = \frac{1}{b^2}
\ee
where $\mu=GM/c^2$ and $b$ has been defined as $h/kc$. Note the dot here is not derivative w.r.t.\ proper time, which is zero for a photon, but w.r.t.\ an affine parameter along the path. We do not need to identify this parameter explicitly, since we are going to work with the {\em shape\/} of the orbit. Specifically, noting $r^2 \dot{\phi} = h$, we have
\be
\left(\frac{\dot{r}^2}{r^4\dot{\phi}^2}\right)^{1/2}=\frac{1}{r^2}\frac{dr}{d\phi} = \left(\frac{1}{b^2}-\frac{1}{r^2}\left(1-\frac{2\mu}{r}\right)\right)^{1/2} \quad
\label{eqn:star}
\ee
Solving (\ref{eqn:star}) for $\mu=0$, which is the limit where the only role of the central mass is to define the origin of the $(r,\phi)$ system), we get $r=b/\sin\phi$. This is the equation of a straight line, which if it continued all the way past the central object, would pass it with perpendicular distance $b$, so we can identify this $b$ as the impact parameter. This generalises what we did above, where we considered the case of a light ray just grazing the surface of an object of radius $R$, for which the unperturbed solution was given by (\ref{eqn:u-straight}), i.e.\ $r=R/\sin\phi$.

Now we consider the exact equation again, and relate to a black hole. We know from the plot of photon effective potential in Fig.~\ref{fig:phot-V-eff}, that photons straying inside $r=3GM/c^2$ are not free to escape to infinity. We therefore want to use this distance as the radius of an `object' which the incoming photon just grazes. Using $R$ for this object radius, we have
\[
\left.\frac{dr}{d\phi}\right|_{r=R} = 0
\]
This tells us that
\[
\frac{1}{b^2}=\frac{1}{R^2}\left(1-\frac{2\mu}{R}\right) \quad \mbox{i.e.} \quad b=R \left(\frac{R}{R-2\mu}\right)^{1/2}
\]
This expression is valid generally, and e.g.\ tells us that the Sun's effective shadow is about 3 km bigger than its coordinate radius, but applied with $R=3GM/c^2$ for a black hole, we find $b=3\sqrt{3}GM/c^2$, which is therefore the radius of the geometric shadow cast. Interestingly, the main effect of including black hole spin in the calculation is to cause a shift in the centroid of the shadow rather than its size --- see e.g.\ \cite{frolov2011introduction} for details.

}

\bigskip

\hrule

\bigskip

For Sgr A* this distance we have just worked out corresponds
to an angular diameter of 52 microarcsec. This is $2.5\times 10^{-10}$
radians, and is very small indeed. Nevertheless, the angular
resolution, approximately equal to the ratio of observing wavelength
to telescope diameter, is about 20 microarcsec for intercontinental mm
VLBI where the effective telescope diameter is 10,000 km.

Interestingly, the 6 billion $\Msun$ black hole in M87 is 1500 times more
massive than Sgr A* and 2000 times further away which makes its event
horizon appear only slightly smaller, at 40 microarcsec. Plans are
also underway to resolve the M87 black hole shadow, together with the
base of its jet which should help us understand how it is accelerated.
All other black holes subtend smaller angular sizes. 3C273, for
example, is 34 times more distant than M87, so its shadow will be
about 1 microarcsec in size.

Near infrared interferometry at ESO's Very Large Telescope in Chile
will use wavelengths around 2 microns and effective diameters of a 100
m to give a resolution of 4 milliarcsec. This will not resolve the
black hole shadow but will resolve the motion of the orbiting stars
much more precisely, particularly as the positional accuracy can be
more than an order of magnitude better than the resolution. General
relativistic effects on the orbits will be seen and fainter, possibly
closer, stars resolved.

Of great interest would be a pulsar orbiting Sgr A*, which if a
millisecond pulsar could prove to be a very stable and reliable
orbiting clock. None have been found yet and the ionized plasma that
bathes the inner region is a problem for sensitive radio
searches.  A  type of pulsar called a magnetar was found in 2013
within 3 arsec of Sgr A*, but this magnetar unfortunately appears to
be rather unreliable as a clock \cite{2014MNRAS.440L..86C}.

A gas cloud, G2, was discovered on a highly eccentric orbit close to
Sgr A* in 2012 \cite{2014arXiv1407.4354P}. At pericentre in 2014, about 3 Earth masses of
material passes within about $2000r_{S}$. Ram pressure from the
accretion flow should cause it to spiral inward over the following few
years leading to an increased accretion rate and associated
display. It is in a similar path to an earlier discovered cloud G1 and
other clouds. Their apocentre and orbital plane lie close to the
orbits of some of the surrounding luminous stars. The winds of the
stars may be the origin of the inspiralling gas clouds.

Further past activity of Sgr A* has been revealed by X-ray fluorescent
emission lines seen from molecular clouds many tens to hundreds of
light years out from the Galactic Centre. One interpretation
suggests that the luminosity of Sgr A* may have been almost a million times
higher about 100 yr ago \cite{2012A&A...545A..35C}.

Sufficient numbers of black hole spin measurements are expected from
X-ray studies of AGN that the spin history can begin to be
understood. Present work indicates many high spin objects in the mass
range from $10^6 - 5\times 10^7\Msun$. The accretion flow within $10\rg$
and particularly within $3\rg$ will be mapped and the corona
understood. Indeed the whole power output of accreting black holes
in terms of radiation, winds and jets will be explored.

Tidal disruption events due to stars straying close to a massive black
hole are beginning to be observed. Black holes with masses above
$10^8\Msun$ can swallow stars whole in the sense that they are not
tidally disrupted until within the event horizon. The tidal forces of
the more common lower-mass black holes can however destroy stars at
greater distances leading to the production of a short lived, very
luminous, accretion disc which decays with a timescale of about a year.

Studies of AGN feedback will accelerate with use of telescopes such as
the Atacama Large Millimtre Array ALMA, the James Webb Space Telescope
JWST, the Large Synoptic Survey Telescope LSST
and the Advanced Telescope for High Energy Astronomy Athena. The
growth of massive black holes through accretion and mergers will
become understood.

Actually testing GR with black holes is difficult. The astrophysical
phenomena observed so far are explained within the Kerr
metric. Some precision tests may occur if a pulsar is serendipitously found
in a close orbit around a stellar mass black hole in our Galaxy or
around Sgr A*. The discovery of gravitational waves from merging black
holes will of course transform this aspect of the study of black
holes.

More generally, the detection of gravitational waves emitted in events involving black holes has great potential for tying down aspects of astrophysical black holes that are poorly understood at present. These include the numbers of intermediate mass holes in the range between stellar mass black holes and SMBHs in galactic centres, and the evolution of the number density of black holes with cosmic epoch. Both these aspects are key targets for a future space-borne gravitational wave detector mission, such as eLISA or a future variant, and form a key component of its science mission \cite{2013arXiv1305.5720C}. There are many further aspects to the importance of gravitational waves in black hole astrophysics, but we defer coverage of these to other chapters of this book.

\bigskip
\hrule
\bigskip

{\it

\section{More general spherically symmetric black holes}

Real black holes are embedded in a universe in which we know there are effects matching those of a cosmological constant, accounting for about 0.7 of the total energy density. Thus it is of interest to embed a black hole in a de Sitter universe. Additionally, black holes in anti-de Sitter universes are very interesting objects theoretically, since via the AdS/CFT correspondence (see e.g.\ \cite{Maldacena:1997re}), the strong gravity effects of black holes can be replaced by weak perturbative effects in a conformal field theory living on the boundary of an anti-de Sitter space (though we note that the dimensionality involved does not necessarily match that of 4d spacetime).

Returning to the derivations of Section~\ref{sect:s-metric}, we find that to satisfy the new field equation
\be
G_{\mu\nu} + \Lambda g_{\mu\nu}= -8 \pi T_{\mu\nu}
\ee
arising from introducing a cosmological constant $\Lambda$,
the new equations the Ricci tensor has to satisfy are
\be
R_{\mu\nu} = \Lambda g_{\mu\nu}
\ee
Using the components already given for the Ricci in (\ref{eqn:Ric-22}) and above, one finds that this leads to a simple modification to the Schwarzschild values of $A$ and $B$. We find that $B$ is still equal to the inverse of $A$, and that now
\be
A = 1-\frac{2GM}{rc^2} -\frac{\Lambda}{3}r^2
\label{eqn:SdS-met}
\ee
These are therefore the metric coefficients for a black hole of mass $M$ embedded in a de Sitter universe, if $\Lambda$ is positive, and anti-de Sitter, if $\Lambda$ is negative.

Horizons occur (for our current form of the metric) when $B \rightarrow \infty$, hence here where $A=0$. If $\Lambda>0$, this now leads to an additional horizon being present, beyond the Schwarzschild one near the black hole centre, at a position close to $r=\sqrt{3/\Lambda}$ (we are adopting a convention here in which $\Lambda$ has dimensions length$^{-2}$). This is a version of the `de Sitter horizon' --- its exact position is modified slightly by the presence of the mass.

A further generalisation we can make, is to the case where the black hole is charged. As mentioned earlier, this is less compelling physically, since we do not expect significant charge separation to have occurred in the formation of a black hole, but the solution for this type of black hole, called the Reissner-Nordstrom solution (see e.g.\ \cite{Wald:1984rg} for references), is still of great interest theoretically. This also falls into our general scheme, and in fact the most general spherically symmetric black hole, which we can dub RNdS, for Reissner-Nordstrom-de Sitter, is given by
\be
B=A^{-1}, \quad A = 1-\frac{2GM}{rc^2} -\frac{\Lambda}{3}r^2 + \frac{q^2}{r^2}
\ee
where $q$ is the black hole charge. Note there are now (in general) {\em three} horizons. The Reissner-Nordstrom solution on its own introduces two horizons --- one like the normal Schwarzschild one, and another usually much smaller one associated with the charge --- and then for $\Lambda>0$ there is an additional de Sitter horizon.

An advantage of dealing with all these cases via the $A$, $B$ form of the metric, is that we can give a unified treatment of energies and angular momenta in orbits. Repeating the analysis of Section~\ref{sect:rel-astro-em} for this new case,
we already have $\dot{r}$ from (\ref{eq-for-rdot}) and can get $\ddot{r}$ from
\[ \frac{d}{dr}\left(\dot{r}^2\right)=\frac{d}{d\tau}\left(\dot{r}^2\right)\frac{d\tau}{dr}
=\frac{2\ddot{r}\dot{r}}{\dot{r}}=2\ddot{r}\]

For a circular orbit, both $\dot{r}$ and $\ddot{r}$ have to vanish
and this combination of expressions therefore says that
\[
-\frac{k^2}{A^2}A' + \frac{2h^2}{r^3} = 0
\]
where $A'\equiv dA/dr$. Combining again with the expression for $\dot{r}=0$ we can solve for $k^2$ and $h^2$, obtaining
\[ k^2 = \frac{2A^2}{2A-rA'}, \quad \text{and} \quad h^2 = \frac{r^3A'}{2A-rA'} \]

So we now have the energy and angular momentum for a particle in a circular orbit for quite a wide range of metrics. To calculate criteria for stability, one can show that it is useful to work in terms of the quantity (see Lasenby (2014) in preparation)
\be
T(r) = \left(\frac{v}{c}\right)^2 =\tanh^2 u
\ee
where $v$ is the velocity in a circular orbit at radius $r$ and $u$ is the corresponding rapidity parameter. From what we have already derived, one can show
\be
T = \frac{r A'}{2A} \quad \text{and} \quad r\dot{\phi} = \gamma v = \sinh u
\ee
and then the criterion for stability turns out to be that
\be
2 T^2 - r T' -2 T
\ee
should be negative. This simple criterion applies across all the spherically symmetric black hole cases. For example, inserting the $A$ appropriate for a Schwarzschild-de Sitter metric, (\ref{eqn:SdS-met}), we find the stability criterion
\be
18 M^2-15 M \Lambda r^3-3 M r+4 \Lambda r^4 <0
\ee
for stable circular orbits (in units with $c=G=1$). For appropriate astrophysical values of $M$ and $\Lambda$, and restoring units, the outer solution to this is well approximated by
\be
r_{\rm stab}\approx \left(\frac{3GM}{4\Lambda c^2}\right)^{1/3}
\ee
We can apply this formula for any central mass, not just a black hole, and one can show that e.g.\ for a cluster of galaxies with $M=10^{15} M_{\odot}$, and for the measured value of the cosmological constant, then $r_{\rm stab} \approx 7.2 {\rm \, Mpc}$. This is interestingly close to the maximum cluster sizes observed, though of course circular orbit stability is not likely to be the direct criterion one would apply in this case, where effects on velocity dispersion would be more appropriate (see \cite{2012MNRAS.422.2945N} for further details).

}

\bigskip
\hrule

\section{Primordial black holes}

Primordial black holes could have formed as a result of the high densities present soon after the Big Bang~\cite{carr10}. Their masses would be comparable to the particle horizon mass at their formation so it can range from the Planck mass ($10^{-8}\kg$) at the Planck time ($10^{-43}\s$), to $10^5\Msun$ at a time of $1\s$. Primordial black holes of mass $10^{12}\kg$ would have formed at $10^{-23}\s$ and would be evaporating now due to the emission of Hawking radiation (see below), which would peak at 100s of MeV. Observations provide a limit on the background intensity of 100 MeV gamma-rays such that evaporating primordial black holes cannot account for more than $10^{-8}$ of the critical density of the Universe~\cite{page76}. The frequency, energies and timescales of the final explosions do not match typical observed gamma-ray bursts. It is therefore unlikely that they are an important constituent of the Universe.

\bigskip
\hrule

{\it

\subsection{Hawking radiation}

An initially very surprising result about black holes was proved by Hawking in 1974 \cite{Hawking:1974rv}. This is that despite their name, the horizon radiates energy, as though it was a blackbody at the temperature of
\be
T=\frac{\hbar c^3}{8\pi k_{\rm B} G M}
\label{eqn:hack-temp}
\end{equation}
where $k_{\rm B}$ is the Boltzmann constant and $M$ is the black hole's mass. The types of black hole we have mainly considered have a minimum mass of several $\msun$, and so their temperatures are less than $\sim 6 \times 10^{-8} {\rm \, K}$, and therefore negligibly tiny. However, a $10^{12}\kg$ primordial black hole of the type mentioned above, would have a temperature of $\sim 1.2 \times 10^{11} {\rm \, K}$, and would therefore produce highly energetic radiation.

The emission process as calculated originally by Hawking was explained in terms of particle production in a second quantised treatment of a massless scalar field surrounding the black hole, but the principles extend to all kinds of field, massive or massless, which will be radiated with `statistics' (i.e.\ Bose-Einstein or Fermi-Dirac) as appropriate to their spin, and with a thermal distribution with the temperature (\ref{eqn:hack-temp}).

A heuristic way of demonstrating this effect is to consider vacuum fluctuations in the space just outside the horizon, which will lead to the production of virtual particle/antiparticle pairs there. Sufficiently close to the horizon, one of the pair will be able to `tunnel' through the horizon and inside can become a real particle with negative energy. (The possibility of negative energy arises since as we saw above (Section~\ref{sect:kerr-part-motion}), the timelike vector in the $t$-direction tips over to become spacelike inside the horizon, which coincides with the Stationary Limit surface in the Schwarzschild case, and so the projection of the particle momentum onto the local $t$-vector may be negative.) The particle remaining outside, which has positive energy, can therefore be emitted off to infinity, whilst the absorption of a negative energy particle by the black hole, decreases the mass $M$, thus providing the overall energy for the emission.

This type of derivation can be used to calculate the energy of the emitted particle, which turns out to be independent of exactly where (though assumed to be close to the horizon) the virtual pair is produced, and yields an energy $E=k_{\rm B}T$ which is only a factor of $2\pi$ higher than the correct result from (\ref{eqn:hack-temp}) --- see Chapter~11 of \cite{2006gere.book.....H} for details.

A further possibility for a calculation avoiding the full rigours of quantum field theory (QFT), is to work with a first-quantised field surrounding the black hole. For example, a detailed derivation for a massive Dirac field is given in Section~8 of \cite{1998RSPTA.356..487L}. This yields precisely the result (\ref{eqn:hack-temp}) together with the correct Fermi-Dirac statistics for a spin-1/2 particle, and the calculation can be extended to electromagnetic and scalar particles, with the same results that would be obtained from QFT in each case (Lasenby {\em et al} unpublished). However, it is only in a QFT context that one can be properly certain of the physics involved, and of how to go about making unique choices for the branches of analytic functions that are involved, and so a QFT approach is still necessary to be fully confident of the results.

\subsection{Link with surface gravity}

What a full QFT approach leads to is a very general result that singles out the `surface gravity' of a black hole (of any type) as the important quantity. The surface gravity would appear to be infinite at the horizon, since no particle can escape the black hole's pull there. However, the definition of surface gravity intended here is that appropriate to the force felt at the horizon, but which would be evaluated by an observer at infinity. In e.g.\ \cite{Wald:1984rg}, Exercise 6.4, this force is shown to differ from that at the horizon by the redshift factor from the horizon to infinity, which is also infinite. Specifically, one finds for a Schwarzschild black hole that the local force necessary to keep an observer of mass $m$ stationary at radial distance $r$ is (see e.g.\ Section~7 of \cite{1998RSPTA.356..487L})
\be
F = \frac{GMm}{r^2}\left(1-\frac{2GM}{rc^2}\right)^{-1/2}
\end{equation}
The gravitational redshift factor between the horizon and infinity is $\left(1-\frac{2GM}{rc^2}\right)^{1/2}$, hence multiplying by this, we obtain the simple expression $GMm/r^2$ as the force that the observer at infinity would ascribe as being necessary to keep the observer at radius $r$ at rest. (In the exercise in
Wald (\cite{Wald:1984rg}), this difference is explained in terms of the different tensions necessary at the ends of a (massless) rope that extends from the observer at infinity to the one at radius $r$.)

This may seem a mundane and wholly Newtonian result, and exactly what we would expect `at ininity'. However, we are now entitled to put $r=2GM/c^2$, i.e.\ the value of $r$ at the horizon, into this expression, in order to evaluate the `surface gravity' of the black hole. This is defined as acceleration per unit mass, and customarily denoted $\kappa$, so we have found
\be
\kappa = \frac{c^4}{4GM}
\label{eqn:S-kappa}
\end{equation}
The contribution of QFT at this point, is to tell us that a horizon with surface gravity $\kappa$ radiates particles with a thermal temperature
\be
T=\frac{\hbar\kappa}{2\pi k_{\rm B}c}
\label{eqn:T-from-kappa}
\end{equation}
Unruh in 1976 \cite{Unruh:1976db} showed that this is the temperature of a thermal heat bath that develops around an observer accelerating at rate $\kappa$ in Minkowski space. We can now understand the link with surface gravity in black holes, at least heuristically. By the equivalence principle and working close to the horizon so that global curvature effects are not important, we can expect that the heat bath seen by a stationary observer feeling the effects of the black hole surface gravity $\kappa$, should be the same as the heat bath seen by an observer in Minkowski space accelerating at the same rate, and we can see indeed that equations (\ref{eqn:S-kappa}) and (\ref{eqn:T-from-kappa}) agree with (\ref{eqn:hack-temp}) for this Schwarzschild case. The advantage of this route, however, is that if we work out the surface gravity in more complicated cases, such as for Kerr or Kerr-Newman or even de Sitter spacetimes (which as we have seen, also possess a horizon), then we can make the same transition to temperature as via (\ref{eqn:T-from-kappa}). For example, for our general spherically symmetric metric (\ref{eqn:gen-metric}), it is easy to show that the surface gravity is
\be
\kappa = \left.\frac{A'}{2\sqrt{AB}}\right|_{B^{-1}=0}
\end{equation}
(The point where $1/B=0$ is picked out as being the position of the event horizon.) Evaluating this for the Reissner-Nordstrom metric discussed above, and for clarity temporarily putting $G=c=1$ etc., so $1/B=A=1-2M/r+q^2/r^2$, we find that the surface gravity at the outer horizon is
\be
\kappa_{RN}=\frac{\sqrt{M^2-q^2}}{\left(M+\sqrt{M^2-q^2}\right)^2}
\end{equation}
and the Hawking temperature of the radiation emitted in this case is given by inserting this $\kappa$ into (\ref{eqn:T-from-kappa}).

\subsection{Astrophysical aspects of black hole evaporation}

We can compute the rate at which a Schwarzschild black hole loses mass via the formula for the luminosity of a blackbody at temperature $T$, i.e.\ $\sigma T^4$ where $\sigma=\pi^2 k_{\rm B}^4/(60\hbar^3 c^2)$ is the Stefan-Boltzmann constant, together with the area of the horizon, which is $4\pi (2GM/c^2)^2$. This yields
\be
\dot{M}c^2 = -\text{power radiated} = -\frac{c^6 \hbar}{15360 \, \pi G^2 M^2}
\label{eqn:power-rad}
\end{equation}
Solving this differential equation, we find that the cube of the mass declines linearly with time:
\be
M^3(t)=M_0^3-\frac{c^4\hbar}{5120\,\pi G^2}t
\end{equation}
where $M_0$ is the initial mass and $t$ is measured from formation. This gives an evaporation time in terms of $M_0$ of
\be
t_{\rm evap} = 2.66\times 10^{-24} M_0^3 {\rm \, yrs}
\end{equation}
If the current age of the universe is $13.8$ billion years, this means only black holes with a mass less than $1.7 \times 10^{11} {\rm \, kg}$ will have had a chance to decay by now. Due to the $M^{-2}$ dependence in the expression for radiated power, (\ref{eqn:power-rad}), nearly all the energy emitted is confined to the very last moments of the black hole's life.

\subsection{Black hole entropy}

The fact that black holes radiate like black bodies, suggest that they should have an entropy. If we identify $Mc^2$ as the hole's energy, then the thermodynamic relation
\be
dU = T dS
\label{eqn:therm-rel}
\end{equation}
along with our identification of the BH temperature in (\ref{eqn:hack-temp}), yields (first working in units with $G=\hbar=c=k_{\rm B}=1$ for clarity)
\be
dM \times 8 \pi M = dS, \quad \text{so that} \quad S= 4 \pi M^2 = \frac{\cla}{4}
\ee
where $\cla=16\pi M^2$ is the black hole area. Putting back the units, we find that the black hole entropy in units of $k_{\rm B}$, is
\be
\frac{S}{k_{\rm B}} = \frac{\cla}{4 \ell_p^2}
\ee
where $\ell_P=\sqrt{\hbar G/c^3}$ is the Planck length and $\cla=4\pi(2GM/c^2)^2$ is the BH area. For an astrophysical black hole, this entropy is large, and as we have seen scales like $M^2$. The result of this is that it is thought that supermassive black holes in the centres of galaxies dominate the entropy budget of the observable universe. Estimates of this are given in \cite{2010ApJ...710.1825E}, with the result that SMBH may contribute up to 7 orders of magnitude more entropy than stellar mass BHs, and approximately 15 orders of magnitude more than the next largest component, due to photons.

This identification of entropy with 1/4 of the event horizon area expressed in units of $\ell_p^2$, is the starting point for a great deal of work connected with whether a microphysics approach to black hole entropy is possible. In this approach we would seek to obtain the same result by counting states, and then assigning $S=k_{\rm B} \ln N$, where $N$ is the number of available microstates. A related question is what happens to the information about the matter and radiation from which the hole was originally formed. This information is screened beyond the horizon, and our lack of knowledge of the internal arrangements and compositions of this original material provides a satisfactory understanding of why the black hole should have an entropy at all, as already proposed by Beckenstein \cite{Bekenstein:1972tm}, {\em before} Hawking's discovery of black hole radiation. However, at the end stage of evaporation of the black hole, we no longer have a screen, and unless the information is somehow encoded in the phases of the emitted particles and fields composing the Hawking radiation, so that this is not random after all, then the information has been irretrievably lost. This would seem to violate the basic principles of quantum mechanics, which demand {\em unitarity} of transformations between beginning and end states, and would not allow the disappearance of information in this way, given that there is no longer a casual horizon beyond which the information is hidden.

These are deep and fundamental questions, which we do not venture further with, but are discussed elsewhere in this volume. We note that our discussion so far has been just in terms of a Schwarzschild black hole, but equivalences of the same nature work for all types of black hole, and are encoded in the `Laws of black hole thermodynamics'. We now briefly discuss these concentrating on their astrophysical implications, and in particular the possibility of extracting energy from rotating black holes.

\subsection{Laws of black hole thermodynamics and the Penrose process}

\label{sect:therm-and-penrose}

The generalisation of (\ref{eqn:therm-rel}) to an uncharged rotating black hole reads (in natural units again, for clarity):
\be
\begin{aligned}
dU &= TdS + \Omega_H dJ \\
\text{i.e.} \quad dM&= \frac{\kappa}{8\pi} d\cla + \Omega_H dJ
\end{aligned}
\label{eqn:bh-1st-law}
\ee
where $\Omega_H$ is the black hole angular velocity at the horizon, and we have again made the identification of temperature with surface gravity/$2\pi$ and of entropy with one quarter of the horizon area. The new feature here is that work can be done to change the angular momentum $J$ of the black hole. (Recall that this is related to the mass and spin parameters via $J=aM$.)

Equation (\ref{eqn:bh-1st-law}) is known as the `First law of black hole thermodynamics'. There are equivalent versions for black holes of the other three laws of thermodynamics as well --- see e.g.\ Section~9.9 of \cite{frolov2011introduction} for a discussion of these. For astrophysical processes, the other law of immediate interest is the equivalent of the Second law, which states that in any classical process, the area of a black hole horizon, which we know measures the black hole's entropy, does not decrease. The restriction to classical processes is necessary, since as we have seen, Hawking radiation succeeds in reducing both mass and surface area.

The most interesting immediate application of these laws is to (classical) processes in which we attempt to remove energy from the hole. This cannot be done for a non-rotating hole, but the $\Omega_H dJ$ term in the First law provides a route through to this for rotating holes.

The first process of this kind to be discussed, was the Penrose process \cite{1969NCimR...1..252P}. In this process, an incoming particle enters the ergosphere of the black hole. It then decays into two particles. Particles can escape from the ergosphere (which lies outside the horizon), and we arrange the initial trajectory and decay such that one of the decay products escapes, and the other falls into the black hole. The key observation is now that inside the ergosphere the $g_t$ Killing vector corresponding to invariance of the metric under time displacements, becomes spacelike, (remember the definition of the ergosphere is where $g_t^2$ changes sign), and it is possible for a particle to have negative energy when its 4-momentum is projected onto it. If this happens, then from conservation of energy, the energy of the emitted particle will be greater than that of the original particle, and we will have succeeded in extracting energy from the black hole. This energy has come via the BH's absorption of the infalling negative energy particle.

One can also analyse what happens to angular momentum in this process, by considering the projections onto the other Killing vector (corresponding to $\phi$ invariance of the metric) $g_{\phi}$. This reveals (see \cite{zee2013einstein} or \cite{frolov2011introduction} for details) that the hole absorbs negative angular momentum as well as negative energy (meaning the emitted particle enjoys a boost to its angular momentum as well as energy, relative to the incoming one) and that the black hole parameter changes obey the inequality
\be
\delta M \ge \Omega_H \delta J
\label{eqn:MJ-ineq}
\ee
Now, the expressions in the Kerr case for the quantities appearing in the First law (\ref{eqn:bh-1st-law}) are
\be
\begin{aligned}
\text{Horizon area} \quad \cla &= 8 \pi \left(M^2 + \sqrt{M^4-J^2}\right)\\
\text{Horizon angular velocity} \quad \Omega_H &= \frac{J}{2M\left(M^2 + \sqrt{M^4-J^2}\right)}\\
\text{Surface gravity} \quad \kappa &= \frac{\sqrt{M^4-J^2}}{2M\left(M^2 + \sqrt{M^4-J^2}\right)}
\end{aligned}
\ee
(again see \cite{zee2013einstein} or \cite{frolov2011introduction} for details). We thus have that the response of the horizon area to a general change in $M$ and $J$ is
\be
\delta \cla = \frac{\partial \cla}{\partial M}\delta M +\frac{\partial \cla}{\partial J}\delta J = \frac{8\pi J}{\Omega_H\sqrt{M^4-J^2}}\left(\delta M-\Omega_H\delta J\right)
\label{eqn:delta-A}
\ee
From (\ref{eqn:MJ-ineq}), we see that despite the fact that both $M$ and $J$ decrease, the r.h.s.\ here is always {\em positive} in the Penrose process. This is an example of the Second law of black hole thermodynamics in action --- for a classical process, the BH entropy, as measured by the horizon area, must always increase.

\subsection{Adiabatic (reversible) changes}

An interesting case to consider astrophysically, is the limit of gradual changes in mass and angular momentum, which satisfy the thermodynamic notion of {\em adiabaticity}, in particular the changes are reversible. This is possible, for example, for particular versions of the Penrose process. In the black hole context, since horizon area equates to entropy, an adiabatic process will have $\delta\cla=0$.

Solving (\ref{eqn:delta-A}) for $\delta\cla=0$, immediately yields the following differential equation for $M(J)$:
\be
\frac{dM}{dJ}=\Omega_H  = \frac{J}{2M\left(M^2 + \sqrt{M^4-J^2}\right)}
\ee
for which the following (implicit) solution works:
\be
2 M_0^2=M^2(J)+\sqrt{M^4(J)-J^2}
\ee
where $M_0$ is a constant. As an example, suppose we start with a Schwarzschild black hole with mass $M_0$, and gradually feed in material with positive angular momentum. We will be able to do this until an extremal black hole with $J=M^2$ is reached, at which point the mass is $\sqrt{2} M_0$. An interesting point is what happens to the horizon radius $r_{\rm outer}$ in this process. We know the general expression for it (in terms of $M$ and $J=aM$) is
\be
r_{\rm outer} = M+\frac{1}{\sqrt{M}}\left(M^2+\sqrt{M^4-J^2}\right)
\ee
and naively we might expect this to increase as the hole gains mass. But in fact we can see for the adiabatic process just discussed,
\be
r_{\rm outer}=\frac{\cla}{8\pi M} = \frac{16 \pi M_0^2}{8\pi M} = \frac{2 M_0^2}{M}
\ee

The horizon radius therefore goes {\em down}, reaching a minimum of $\sqrt{2} M_0$ at the end of the process (corresponding to the $r_{\rm outer}=M$ value for an extremal black hole), having started at $2 M_0$.

\subsection{Other processes for extracting energy from a spinning black hole}

\label{sect:superrad}

As well as the Penrose process, there are other ways of extracting energy from a rotating black hole. One of these is `superradiance', which is the analogue for waves of what happens for particles in the Penrose process. The possibility of such a process was first drawn attention to specifically in the context of the Kerr solution by Starobinski in 1973 \cite{1973JETP...37...28S}. It involves an incoming radiation field of the form
\be
\phi \sim \phi_0(r,\theta) e^{-i\omega t} e^{m\phi}
\ee
for which part of the wave (the `transmitted' wave) is absorbed by the black hole, and the other part (the `reflected' wave) reaches infinity again. In the same way as for the Penrose process, due to the spacelike nature of the $g_t$ Killing vector inside the ergosphere, the transmitted wave can have negative energy, so that the reflected wave carries increased energy to infinity (specifically it will have greater amplitude at the same frequency as compared to the incident wave). We can derive the condition for this to occur (at least in a heuristic way), as follows (see e.g.\ Section 8.8 of \cite{frolov2011introduction}).

We can think of the wave as composed of quanta with energy $\hbar \omega$ and angular momentum $\hbar m$. When these are absorbed by the black hole, the change in the black hole's parameters will satisfy
\be
\frac{\delta J}{\delta M} = \frac{\hbar m}{\hbar \omega}
\ee
But we also know that $\delta M - \Omega_H\delta J\ge 0$ from the fact the horizon area must not decrease. Putting these together leads to
\be
\frac{\delta M}{\omega}\left(\omega-\Omega_H m\right) \ge 0
\ee
and so the condition for superradiance ($\delta M <0$) is
\be
0 < \omega < m \Omega_H
\ee
Further details of this process, which only works for bosonic fields, can be found in Section~12.4 of \cite{Wald:1984rg}.

Our final example, concerns what may be an important way in which spinning black holes return energy to the environment, and which was referred to briefly above in Section~\ref{sect:jets}, namely the Blandford-Znajek effect \cite{Blandford:1977ds}.

Here, one considers a Kerr black hole immersed in an ambient magnetic field, typically associated with an accretion disc around the black hole. A possible picture of the process (see e.g.\ \cite{frolov2011introduction} Section~8.9) represents the black hole horizon as a moving conductor within the magnetic field. Indeed the Kerr black hole horizon, in common with all stationary event horizons, can be modelled as having a resistance of $4\pi$ in geometrical units (377 Ohms in ordinary units). This rotating conductor coupled to the magnetic field, generates a current which flows between the poles and equator --- it is in this sense like a dynamo. The main analytic field configuration chosen by Blandford \& Znajek was a `split monopole' --- i.e.\ a different sign magnetic monopole solution in each hemisphere, with magnetic field lines pointing radially outwards or inwards. This needs a current sheet on the join between them in the equitorial plane, which is presumed to be supplied at least outside the hole by currents in the accretion disc.

The radiated power can be calculated from the Poynting vector, and gives  similar results as for a rotating magnetic dipole pulsar model evaluated on the light cylinder. Specifically we find (see e.g.\ \cite{meier2009black} Section~14.3)
\be
\begin{aligned}
\mathcal{P^{EM}} &= \frac{B_E^2 r_E^4 \Omega_f^2}{c} = \frac{B_H^2 r_H^4 \Omega_f^2}{c}\\
&\sim 4.1 \times 10^{47} {\rm \, erg \, s^{-1}} \left(\frac{B_H}{10^5 {\rm \, G}}\right)^2 \left(\frac{a}{M}\right)^2 \left(\frac{M}{10^9\msun}\right)^2
\end{aligned}
\ee
Here $B_E$ and $B_H$ are the magnetic field values at the ergosphere and horizon respectively, $r_E$ and $r_H$ the corresponding $r$ values, and $\Omega_f$ the magnetic field rotation rate, which is typically $\sim \frac{1}{2} \Omega_H$. This EM power will presumably be manifested in a magnetised particle wind and jet, removing angular momentum and energy from the black hole in the process. Subsequent work (see e.g.\ Komissarov \cite{2004MNRAS.350..427K}) has shown that the Blandford-Znajek (BZ) mechanism is stable, and clarified that important components of it take place within the ergosphere itself, emphasising its links with the Penrose process and superradiance. The BZ mechanism is now thought to be an important component in generating at least some of the high energy jets we see in astrophysics \cite{2011MNRAS.418L..79T}.

}

\bigskip
\hrule
\bigskip

\section{Conclusions}

Black holes are now an integral features of our cosmic landscape. Many
millions of stellar mass black holes reside in our Galaxy. A
supermassive black hole lies at the centre of all massive galaxies.
We know little about most of them, unless they are massive and nearby
or accreting gas from a stellar companion or the surrounding
interstellar medium. The inner part of the accretion flow can be
extremely luminous. making the immediate surroundings of the blackest
parts of the Universe into the brightest.

The behaviour of observed black holes can so far be explained well by
General Relativity and the Kerr metric. The physics of accretion onto
black holes, and of the outflows and jets that often accompany
inflows, is complex and leads to complicated observational phenomena
which lie at the forefront of astrophysics. Understanding how quasars
work remains a significant astrophysical challenge. The consequences
for a galaxy hosting a supermassive black hole are profound. It is
likely that the energy released by the growth of the black hole plays
a decisive factor in its final stellar mass and possibly its physical
size.

Black holes are intrinsically relativistic objects which have
stimulated and tested physical understanding to the extreme. The
internal structure of the black hole, within the event horizon, is
beyond direct observation and we have not discussed it here, although it is clearly of great theoretical interest. In this Chapter we have
looked at black holes as observable physical objects and considered
how they work as engines of gravitational energy release.

The future of research into astrophysical black holes is very bright,
as more telescopes probe ever deeper into more wavebands to uncover
new objects, features and phenomena. Serendipitous discovery plays a
key role in the history of astronomy and we look forward to the
discoveries of a millisecond pulsar orbiting close to a massive black
hole, a rogues gallery of black hole shadows, a range of quasars above
redshift 10, a complete X-ray spectral-timing deconstruction of the
innermost accretion flow around a rapidly spinning black hole and,
most of all, some new phenomena of black holes that we have not yet
anticipated.

\bibliography{book_chap}


\end{document}